\documentclass[twocolumn,eqsecnum,nofootinbib,showpacs,preprintnumbers,widetext]{revtex4}

\usepackage{amsmath,amsfonts,latexsym,amssymb,times}
\allowdisplaybreaks[4]

\newcommand{\rf}[1]{(\ref{#1})}

\def\be{\begin{equation}}
\def\ee{\end{equation}}
\def\beq{\begin{eqnarray}}
\def\eeq{\end{eqnarray}}

\def\parline{\,\partial\kern -0.55em /\,\,}

\def\half{{\frac{1}{2}}}

\def\DD{{\cal D}}

\def\LL{{\cal L}}

\def\TT{{\cal T}}

\def\smzero{{\scriptscriptstyle (0)}}
\def\smone{{\scriptscriptstyle (1)}}

\def\smzero{{\scriptscriptstyle (0)}}
\def\smone{{\scriptscriptstyle (1)}}

\def\sc{{\rm sc}}
\def\cur{{\rm cur}}
\def\sh{{\rm sh}}
\def\eff{{\rm eff}}
\def\st{{\rm st}}

\def\stand{{\rm stand}}
\def\impr{{\rm impr}}

\def\Cbf{{\bf C}}

\begin{document}

\preprint{FIAN-TD-2010-13; arXiv: 1011.4261 [hep-th] }

\title{Gauge invariant approach to low-spin anomalous conformal currents and shadow fields}

\author{ R.R. Metsaev}

\email{metsaev@lpi.ru}

\affiliation{ Department of Theoretical Physics, P.N. Lebedev Physical
Institute, Leninsky prospect 53,  Moscow 119991, Russia}

\begin{abstract}
Conformal low-spin anomalous currents and shadow fields in flat space-time of
dimension greater than or equal to four are studied. Gauge invariant
formulation for such currents and shadow fields is developed. Gauge
symmetries are realized by involving Stueckelberg and auxiliary fields. Gauge invariant
differential constraints for anomalous currents and shadow fields and
realization of global conformal symmetries are obtained. Gauge invariant
two-point vertices for anomalous shadow fields are also obtained. In
Stueckelberg gauge frame, these gauge invariant vertices become the standard
two-point vertices of CFT. Light-cone gauge two-point vertices of the
anomalous shadow fields are derived. AdS/CFT correspondence for anomalous
currents and shadow fields and the respective normalizable and
non-normalizable solutions of massive low-spin AdS fields is studied. The
bulk fields are considered in modified de Donder gauge that leads to
decoupled equations of motion. We demonstrate that leftover on-shell gauge
symmetries of bulk massive fields correspond to gauge symmetries of boundary
anomalous currents and shadow fields, while the modified (Lorentz) de Donder
gauge conditions for bulk massive fields correspond to differential
constraints for boundary anomalous currents and shadow fields.
\end{abstract}


\maketitle

\section{Introduction}

In space-time of dimension $d\geq 4$, fields of CFT can be separated into
two groups: conformal currents and shadow fields. Field having Lorentz
algebra spin $s$ and conformal dimension $\Delta = s+d-2$, is referred to as
conformal current with canonical dimension, while field having Lorentz
algebra spin $s$ and conformal dimension $\Delta > s+d-2$ is referred to as
anomalous conformal current. Accordingly, field having Lorentz algebra spin
$s$ and conformal dimension
$\Delta = 2 - s$, is referred to as shadow field with canonical dimension%
\footnote{ It is the shadow fields having canonical dimension that are used
to discuss conformal invariant equations of motion and Lagrangian
formulations (see e.g.
Refs.\cite{Fradkin:1985am,Segal:2002gd,Erdmenger:1997gy,
Boulanger:2001he,Shaynkman:2004vu,Vasiliev:2009ck}). In earlier literature,
discussion of shadow field dualities may be found in
Ref.\cite{Petkou:1994ad}.},
while field having Lorentz algebra spin $s$ and conformal dimension $\Delta <
2 - s$ is referred to as anomalous shadow field.

In Refs.\cite{Metsaev:2008fs,Metsaev:2009ym}, we developed the gauge
invariant (Stueckelberg) approach to the conformal currents and shadow fields
having canonical conformal dimensions. In the framework of AdS/CFT
correspondence such currents and shadow fields are related to {\it massless}
AdS fields. The purpose of this paper is to develop gauge invariant approach
to the anomalous conformal currents and shadow fields which, in the framework
of AdS/CFT correspondence, are related to {\it massive} AdS fields. The
examples of spin-1 and spin-2 conformal fields demonstrate all characteristic
features of our approach. In this paper, because these examples are very
important in their own right, we discuss spin-1 and spin-2 anomalous
conformal currents and shadow fields. Arbitrary spin anomalous conformal
currents and shadow fields will be considered in forthcoming publication. Our
approach can be summarized as follows.
\\
{\bf i}) Starting with field content of the standard formulation of anomalous
conformal currents (and anomalous shadow fields), we introduce Stueckelberg
fields and auxiliary fields, i.e., we extend space of fields entering the
standard CFT.
\\
{\bf ii}) On the extended space of currents (and shadow fields), we introduce
differential constraints, gauge transformations, and conformal algebra
transformations. These differential constraints are invariant under the gauge
transformations and the conformal algebra transformations.
\\
{\bf iii})  The gauge symmetries and the differential constraints make it
possible to match our approach and the standard one, i.e., by appropriate
gauge fixing to exclude the Stueckelberg fields and by solving differential
constraints to exclude the auxiliary fields we obtain the standard formulation of anomalous conformal currents
and shadow fields.

We apply our approach to the study of AdS/CFT correspondence between
massive AdS fields and corresponding boundary anomalous conformal currents
and shadow fields. We demonstrate that normalizable modes of massive AdS
fields are related to anomalous conformal currents, while non-normalizable
modes of massive AdS fields are related to anomalous shadow fields. In the
earlier literature, the correspondence between non-normalizable bulk modes
and shadow fields was studied in Ref.\cite{Mueck:1998iz} (for spin-1 fields)
and in Ref.\cite{Polishchuk:1999nh} (for spin-2 fields). To our knowledge,
AdS/CFT correspondence between normalizable massive modes and anomalous
conformal currents has not been considered in the earlier literature. As
compared to the studies in Refs.\cite{Mueck:1998iz,Polishchuk:1999nh}, our
approach involves large amount of gauge symmetries. Therefore results of
these references are obtained from the ones in this paper by using some
particular gauge condition, which we refer to as Stueckelberg gauge fixing.
We note also that our approach provides quick access to the light-cone gauge
formulation of CFT. Perhaps, one of the main advantages of our
approach is that this approach gives easy access to the study of AdS/CFT
correspondence in light-cone gauge frame. This is very important for future
application of our approach to studying string/gauge theory dualities because
one expects that string theory in AdS/Ramond-Ramond background can be
quantized only in light-cone gauge.

Our approach to the study of AdS/CFT correspondence can be summarized as follows.
\\
{\bf i}) We use CFT adapted gauge invariant approach to AdS field dynamics
developed in Ref.\cite{Metsaev:2009hp}. For spin-1 and spin-2 massive AdS
fields, we use the respective modified Lorentz gauge and modified de Donder
gauge. Remarkable property of these gauges is that they lead to the simple
{\it decoupled} bulk equations of motion which can be solved in terms of
Bessel function and this simplifies considerably study of AdS/CFT
correspondence. Also, using these gauges, we demonstrate that the two-point
gauge invariant vertex of the anomalous shadow field does indeed emerge from
massive AdS field action when it is evaluated on solution of the Dirichlet
problem. AdS field action evaluated on solution of the Dirichlet problem will
be referred to as effective action in this paper.
\\
{\bf ii}) The number of boundary gauge fields involved in our gauge invariant
approach to the anomalous conformal current (or anomalous shadow field) coincides
with the number of bulk massive gauge AdS fields involved in the standard gauge
invariant Stueckelberg approach to massive field. Note however that, instead
of the standard gauge invariant approach to massive field, we use the CFT adapted
formulation of massive AdS field developed in
Ref.\cite{Metsaev:2009hp}%
\footnote{We note also that the number of gauge transformation parameters
involved in our gauge invariant approach to anomalous current (or anomalous
shadow field) coincides with the number of gauge transformation parameters of
bulk massive gauge AdS field involved in the standard gauge invariant
approach to massive field.}.
\\
{\bf iii}) Our modified Lorentz gauge (for spin-1 massive AdS field) and
modified de Donder gauge (for spin-2 massive AdS field) turn out to be
related to the differential constraints we obtained in the framework of gauge
invariant approach to the anomalous conformal currents and shadow fields.
\\
{\bf iv}) {\it Leftover on-shell} gauge symmetries of massive bulk AdS
fields are related to the gauge symmetries of boundary anomalous conformal
currents (or anomalous shadow fields).

The rest of the paper  is organized as follows.

In Sec. \ref{sec02},  we summarize the notation used in this paper.

In Secs. \ref{sec03} and \ref{sec04}, we  start with  the respective examples
of the spin-1 anomalous conformal current and spin-1 anomalous shadow field.
We illustrate our gauge invariant approach to describing the anomalous
conformal current and shadow field. For the spin-1 anomalous shadow field, we
obtain the gauge invariant two point vertex and discuss how our gauge
invariant approach is related to the standard approach to CFT. Also, using
our gauge invariant approach we obtain light-cone gauge description of the
spin-1 anomalous conformal current and shadow field.

Secs. \ref{sec05} and \ref{sec06} are devoted to spin-2 anomalous conformal
current and spin-2 anomalous shadow field respectively. In these Sections we
generalize results of Secs. \ref{sec03} and \ref{sec04} to the case of spin-2
anomalous conformal current and shadow field.

In Sec. \ref{sec09}, we discuss two-point current-shadow field interaction
vertex.

In Sec. \ref{secAdS/CFT}, because use  of modified Lorentz (de Donder) gauge
makes study of AdS/CFT correspondence for spin-1 (spin-2) field similar to
the one for scalar field, we briefly review the AdS/CFT correspondence
for the scalar field.

Sec. \ref{secAdS/CFTspin1} is devoted to the study of AdS/CFT correspondence
for bulk spin-1 massive AdS field and boundary spin-1 anomalous conformal
current and shadow field, while in Sec.\ref{secAdS/CFTspin2} we extend
results of Sec. \ref{secAdS/CFTspin1} to the case of spin-2 fields.

We collect various technical details in two appendices. In Appendices
\ref{app-01} and \ref{app-02} we present details of the derivation of CFT
adapted gauge invariant Lagrangian for the respective spin-1 and spin-2
massive AdS fields.

\section{Preliminaries}
\label{sec02}

\subsection{Notation}

Our conventions are as follows. $x^a$ denotes coordinates in $d$-dimensional
flat space-time, while $\partial_a$ denotes derivatives with respect to
$x^a$, $\partial_a \equiv \partial /
\partial x^a$. Vector indices of the Lorentz algebra $so(d-1,1)$ take
the values $a,b,c,e=0,1,\ldots ,d-1$. We use mostly positive flat metric
tensor $\eta^{ab}$. To simplify our expressions we drop $\eta_{ab}$ in scalar
products, i.e., we use $X^aY^a \equiv \eta_{ab}X^a Y^b$. Throughout this
paper we use operators constructed out of the derivatives and coordinates,
\be
\Box=\partial^a\partial^a\,,\qquad
x\partial \equiv x^a \partial^a \,, \qquad x^2 = x^a x^a\,.
\ee

Sometimes we use light-cone frame. In the light-cone frame, space-time
coordinates are decomposed as $x^a= x^+, x^-,x^i$, where light-cone
coordinates in $\pm$ directions are defined as $x^\pm=(x^{d-1} \pm
x^0)/\sqrt{2}$ and $x^+$ is taken to be a light-cone time. $so(d-2)$ algebra
vector indices take values $i,j =1,\ldots, d-2$. We adopt the conventions:
\be \partial^i=\partial_i\equiv\partial/\partial x^i\,, \qquad
\partial^\pm=\partial_\mp \equiv
\partial/\partial x^\mp \,.\ee

\subsection{Global conformal symmetries }

In $d$-dimensional flat space-time, the conformal algebra $so(d,2)$ consists
of translation generators $P^a$, dilatation generator $D$, conformal boost
generators $K^a$, and generators of the $so(d-1,1)$ Lorentz algebra $J^{ab}$.
We assume the following normalization for commutators of the conformal
algebra:
\beq
&& {}[D,P^a]=-P^a\,, \hspace{0.5cm}  {}[P^a,J^{bc}]=\eta^{ab}P^c
-\eta^{ac}P^b, \ \ \ \
\nonumber\\
&& [D,K^a]=K^a\,, \hspace{0.7cm}
[K^a,J^{bc}]=\eta^{ab}K^c - \eta^{ac}K^b,\qquad
\nonumber\\[-6pt]
\label{man12112010-01} &&
\\[-6pt]
&& \hspace{1.5cm} {}[P^a,K^b]=\eta^{ab}D - J^{ab}\,,
\nonumber\\
&& \hspace{1.5cm} [J^{ab},J^{ce}]=\eta^{bc}J^{ae}+3\hbox{ terms} \,.
\nonumber
\eeq

Let $\phi$ denotes conformal current (or shadow field) in flat space-time of
dimension $d\geq 4$. Under conformal algebra transformations the $\phi$
transforms as
\be \label{04092008-01} \delta_{\hat{G}} \phi  = \hat{G} \phi \,, \ee
where realization of the conformal algebra generators $\hat{G}$ in
terms of differential operators acting on the $\phi$ takes the form
\beq
\label{conalggenlis01} && P^a = \partial^a \,,
\\[3pt]
\label{conalggenlis02} && J^{ab} = x^a\partial^b -  x^b\partial^a + M^{ab}\,,
\\[3pt]
\label{conalggenlis03} && D = x\partial  + \Delta\,,
\\[3pt]
\label{conalggenlis04} && K^a = K_{\Delta,M}^a + R^a\,,
\\[5pt]
\label{kdelmdef01} && \hspace{1cm} K_{\Delta,M}^a \equiv -\frac{1}{2}x^2\partial^a + x^a D
+ M^{ab}x^b\,.\qquad
\eeq
In \rf{conalggenlis02}-\rf{conalggenlis04}, $\Delta$ is operator of conformal
dimension, $M^{ab}$ is spin operator of the Lorentz algebra. Action of
$M^{ab}$ on fields of the Lorentz algebra is well known and for rank-2 tensor,
vector, and scalar fields considered in this paper is given by
\beq
&& M^{ab}\phi^{ce} = \eta^{ae}\phi^{cb} +\eta^{ac}\phi^{be} -
(a\leftrightarrow b)\,,
\nonumber\\
&& M^{ab}\phi^c = \eta^{ac}\phi^b - (a\leftrightarrow b)\,,
\\
&& M^{ab}\phi = 0\,.
\nonumber
\eeq
These relations imply that action of operator $K_{M,\Delta}^a$ \rf{kdelmdef01} on the fields can be presented as
\beq
&&  K_{\Delta,M}^a\phi^{bc}= K_\Delta^a \phi^{bc} + M^{abf} \phi^{fc} +
M^{acf} \phi^{bf}\,,\qquad
\nonumber\\[5pt]
&& K_{\Delta,M}^a\phi^b = K_\Delta^a \phi^b + M^{abf} \phi^f\,,
\\[5pt]
&& K_{\Delta,M}^a\phi = K_\Delta^a \phi\,,
\nonumber\\[5pt]
\label{man19112010-01} && \hspace{1.5cm} K_\Delta^a \equiv -\half x^2 \partial^a + x^a
(x\partial + \Delta) \,,
\\[5pt]
\label{man19112010-02} && \hspace{1.5cm}  M^{abc} \equiv \eta^{ab}x ^c -\eta^{ac} x^b \,.
\eeq
In \rf{conalggenlis04}, $R^a$ is operator depending, in general, on the derivatives
with respect to the space-time coordinates%
\footnote{For conformal currents and shadow fields studied in this paper, the
operator $R^a$ does not depend on the derivatives. Dependence of
$R^a$ on derivatives appears e.g., in ordinary-derivative approach to conformal fields
\cite{Metsaev:2007fq}.}
and not depending on the space-time coordinates $x^a$. In the standard formulation of
conformal currents and shadow fields, the operator $R^a$ is equal to zero,
while in the gauge invariant approach that we develop in this paper, the operator
$R^a$ is non-trivial. This implies that, in the framework of the gauge invariant
approach, the complete description of the conformal currents and shadow fields
requires, among other things, finding the operator $R^a$.

\section{ Spin-1 anomalous conformal current}
\label{sec03}

In this section, we develop gauge invariant approach to spin-1 anomalous
conformal current. Besides the gauge invariant formulation we discuss two
gauge conditions which can be used for studying the anomalous conformal
currents - Stueckelberg gauge and light-cone gauge. We would like to discuss
these gauges because of the following reasons.

\noindent {\bf i}) It turns out that the Stueckelberg gauge reduces our
approach to the standard formulation of CFT. Therefore the use of the
Stueckelberg gauge allows us to demonstrate how the standard approach to
anomalous conformal currents is obtained from our gauge invariant approach.

\noindent {\bf ii}) Motivation for considering the light-cone gauge frame
cames from conjectured duality of the SYM theory and the theory of the
superstring in AdS background \cite{Maldacena:1997re}. By analogy with flat
space, we expect that a quantization of the Green-Schwarz AdS superstring
\cite{Metsaev:1998it} will be straightforward only in the light-cone gauge
\cite{Metsaev:2000yf,Metsaev:2000yu}. Therefore it seems that from the
stringy perspective of AdS/CFT correspondence, light-cone approach to CFT
is the fruitful direction to go.

\subsection{ Gauge invariant formulation }
\label{sec03G}

To discuss gauge invariant formulation of spin-1 anomalous conformal current
in flat space of dimension $d\geq 4$ we use one vector field
$\phi_{\cur,0}^a$ and two scalar fields $\phi_{\cur,1}$, $\phi_{\cur,-1}$:
\be \label{man24102010-01}  \phi_{\cur,0}^a\,, \qquad \phi_{\cur,-1}\,, \qquad
\phi_{\cur,1}\,. \ee
The fields $\phi_{\cur,0}^a$ and $\phi_{\cur,\pm1}$ transform in the
respective vector and scalar irreps of the Lorentz algebra $so(d-1,1)$. We
note that fields \rf{man24102010-01} have the conformal dimensions
\be \label{man24102010-02} \Delta_{\phi_{\cur,0}^a} = \frac{d}{2} + \kappa\,, \qquad
\Delta_{\phi_{\cur,\pm 1}} = \frac{d}{2} + \kappa \pm 1 \,, \ee
where $\kappa$ is a dimensionless parameter. In the framework of AdS/CFT
correspondence, $\kappa$ is related to the mass parameter $m$ of spin-1
massive AdS field as
\be \label{man24102010-02n}
\kappa \equiv \sqrt{m^2 + \frac{(d-2)^2}{4}}\,.\ee

We now introduce the following differential constraint:
\beq
\label{man24102010-03}
&& \partial^a \phi_{\cur,0}^a  + r_z^{00}\Box\phi_{\cur,-1} + r_\zeta^{00} \phi_{\cur,1} =
0 \,,
\\[5pt]
&& \hspace{1cm}  r_z^{00} \equiv
\Bigl(\frac{2\kappa+d-2}{4\kappa}\Bigr)^{1/2}\,,
\nonumber\\[-5pt]
\label{man24102010-04} &&
\\[-5pt]
&& \hspace{1cm} r_\zeta^{00} \equiv
\Bigl(\frac{2\kappa-d+2}{4\kappa}\Bigr)^{1/2}\,.
\nonumber
\eeq

One can make sure that this constraint is invariant under the gauge
transformations
\beq \label{man24102010-06}
&& \delta\phi_{\cur,0}^a = \partial^a \xi_{\cur,0}
\\[5pt]
\label{man24102010-07} && \delta\phi_{\cur,-1} = - r_z^{00}\xi_{\cur,0}\,,
\\[5pt]
\label{man24102010-08} && \delta\phi_{\cur,1} = - r_\zeta^{00} \Box \xi_{\cur,0}\,,
\eeq
where $\xi_{\cur,0}$ is a gauge transformation parameter.

To complete our gauge invariant formulation  we provide realization of the
operator $R^a$ on space of gauge fields \rf{man24102010-01},
\beq
&& R^a \phi_{\cur,0}^b =  - 2\kappa r_z^{00} \eta^{ab}\phi_{\cur,-1}^{}\,,
\nonumber\\[5pt]
\label{man24102010-08a} && R^a \phi_{\cur,-1} = 0\,,
\\[5pt]
&& R^a \phi_{\cur,1}=  - 2\kappa r_\zeta^{00} \phi_{\cur,0}^a \,.
\nonumber
\eeq
Using \rf{man24102010-08a}, we make sure that constraint \rf{man24102010-03}
is invariant under transformations of the conformal algebra \rf{04092008-01}.

\subsection{ Stueckelberg gauge frame }
\label{sec03S}

We now discuss the spin-1 anomalous conformal current in the Stueckelberg
gauge frame. From \rf{man24102010-07}, we see that the scalar field
$\phi_{\cur,-1}$ transforms as Stueckelberg field, i.e., this field can be
gauged away via Stueckelberg gauge fixing,
\be \label{man24102010-10n} \phi_{\cur,-1}=0.\ee
Using this gauge in constraint \rf{man24102010-03}, we see that the remaining
scalar field $\phi_{\cur,1}$ can be expressed in terms of the vector field
$\phi_{\cur,0}^a$,
\be \label{man24102010-10}
\phi_{\cur,1}  = -\frac{1}{ r_\zeta^{00} }\partial^a \phi_{\cur,0}^a \,,
\ee
i.e., making use of the gauge symmetry and differential constraint
\rf{man24102010-03} we reduce field content of our approach
\rf{man24102010-01} to the one in the standard approach. In other words, the
gauge symmetry and differential constraint make it possible to match our
approach and the standard formulation of spin-1 anomalous
conformal current%
\footnote{ As in standard approach to CFT, our currents can be considered
either as fundamental field degrees of freedom or as composite operators. At
the group theoretical level, we study in this paper, this distinction is
immaterial. Discussion of interesting methods for building conformal currents
as composite operators may be found in
Refs.\cite{Konstein:2000bi,Gelfond:2006be}.}.

\subsection{ Light-cone gauge frame }

For the spin-1 anomalous conformal current, the light-cone gauge frame is
achieved through the use of differential constraint \rf{man24102010-03} and
light-cone gauge condition. Using gauge symmetry of the spin-1 anomalous
conformal current \rf{man24102010-06}, we impose the light-cone gauge on the
field $\phi_{\cur,0}^a$,
\be \label{man05112010-29} \phi_{\cur,0}^+ =0 \,.\ee
Using this gauge in differential constraint \rf{man24102010-03}, we
find
\be  \label{man05112010-30}  \phi_{\cur,0}^- = -
\frac{\partial^j}{\partial^+}\phi_{\cur,0}^j -
\frac{r_z^{00}}{\partial^+}\Box\phi_{\cur,-1} -
\frac{r_\zeta^{00}}{\partial^+}\phi_{\cur,1} \,. \ee
We see that we are left with vector field $\phi_{\cur,0}^i$ and two scalar
fields $\phi_{\cur,\pm1}$. These fields constitute the field content of the
light-cone gauge frame.

\section{ Spin-1 anomalous shadow field}
\label{sec04}

\subsection{ Gauge invariant formulation }

To discuss gauge invariant formulation of spin-1 anomalous shadow field in space of
dimension $d\geq 4$ we use one vector field $\phi_{\sh,0}^a$ and two scalar
fields $\phi_{\sh,-1}$, $\phi_{\sh,1}$:
\be  \label{man24102010-20} \phi_{\sh,0}^a\,, \qquad \phi_{\sh,-1}\,,\qquad
\phi_{\sh,1}\,. \ee
The fields $\phi_{\sh,0}^a$ and $\phi_{\sh,\pm 1}$ transform in the respective
vector and scalar representations of the Lorentz algebra $so(d-1,1)$. We note
that these fields have the conformal dimensions
\be \label{man24102010-21} \Delta_{\phi_{\sh,0}^a} = \frac{d}{2} - \kappa\,,
\qquad \Delta_{\phi_{\sh,\pm 1}} = \frac{d}{2} - \kappa \pm 1\,. \ee
In the framework of AdS/CFT correspondence, $\kappa$ is related to the mass
parameter $m$ of spin-1 massive AdS field as in \rf{man24102010-02n}.

We now introduce the following differential constraint:
\be \label{man24102010-22}
\partial^a \phi_{\sh,0}^a  + r_\zeta^{00}\Box\phi_{\sh,-1} + r_z^{00}\phi_{\sh,1}
= 0 \,,
\ee
where $r_z^{00}$, $r_\zeta^{00}$ are given in
\rf{man24102010-04}. We make sure that constraint
\rf{man24102010-22} is invariant under the gauge transformations
\beq \label{man24102010-23}
&& \delta\phi_{\sh,0}^a = \partial^a \xi_{\sh,0}
\\[5pt]
\label{man24102010-24} && \delta\phi_{\sh,-1} = - r_\zeta^{00} \xi_{\sh,0}\,.
\\[5pt]
\label{man24102010-25} && \delta\phi_{\sh,1} = - r_z^{00}\Box\xi_{\sh,0}\,,
\eeq
where $\xi_{\sh,0}$ is a gauge transformation parameter.

To complete our gauge invariant formulation  of the spin-1 anomalous shadow field
we provide realization of the operator $R^a$ on space of gauge fields
\rf{man24102010-20},
\beq
&& R^a \phi_{\sh,0}^b = 2\kappa r_\zeta^{00} \eta^{ab}\phi_{\sh,-1}^{}\,,
\nonumber\\[5pt]
\label{man24102010-26n} && R^a \phi_{\sh,-1} = 0\,,
\\[5pt]
&& R^a \phi_{\sh,1}= 2\kappa r_z^{00} \phi_{\sh,0}^a
\,.
\nonumber
\eeq

We proceed with the discussion of two-point vertex for the
spin-1 anomalous shadow field. The gauge invariant two-point vertex we find takes the
form
\beq \label{man05112010-18}
\Gamma \ \ & = & \int d^dx_1 d^dx_2 \Gamma_{12} \,,
\\[5pt]
\label{man05112010-19}\Gamma_{12}\! & = &  \frac{\phi_{\sh,0}^a(x_1)
\phi_{\sh,0}^a(x_2)}{2|x_{12}|^{2\kappa + d}}
\nonumber\\[5pt]
& +  & \sum_{\lambda=\pm
1}\frac{\omega_\lambda}{2|x_{12}|^{2\kappa+d-2\lambda}}
\phi_{\sh,\lambda}(x_1)\phi_{\sh,\lambda}(x_2)\,,\qquad
\\
&& \omega_1 = \frac{1}{2\kappa(2\kappa+d-2)}\,,
\nonumber\\[-5pt]
\label{man05112010-19b} &&
\\[-5pt]
&& \omega_{-1} =  2(\kappa+1)(2\kappa+d)\,,
\nonumber
\eeq
\be \label{man05112010-20}
|x_{12}|^2 \equiv x_{12}^a x_{12}^a\,, \qquad x_{12}^a = x_1^a - x_2^a\,.
\ee
One can check that this vertex is invariant under the gauge transformations
of the spin-1 anomalous shadow field given in
\rf{man24102010-23}-\rf{man24102010-25}. Also, we check that the vertex is invariant
under the conformal algebra transformations.

The kernel of the vertex $\Gamma$ is related to a two-point correlation
function of the spin-1 anomalous conformal current. In our approach, the spin-1
anomalous conformal current is described by gauge fields given in \rf{man24102010-01}.
Therefore, in order to discuss the correlation function of the anomalous conformal
current in a proper way, we should impose a gauge condition on
the gauge fields in \rf{man24102010-01}.%
\footnote{ We note that, in the gauge invariant approach, correlation
functions of the conformal current can be studied without gauge fixing. To do
that one needs to construct gauge invariant field strengths for the gauge
potentials $\phi_{\cur,0}^a$, $\phi_{\cur,\pm 1}$. Study of field strengths
for the conformal current is beyond the scope of this paper. Recent
interesting discussion of method for building field strengths may be found in
Refs.\cite{Bastianelli:2008nm,Corradini:2010ia}.}
We have considered the spin-1 anomalous conformal current in the Stueckelberg
and light-cone gauge frames. This is to say that correlation function of the
spin-1 anomalous conformal current in the Stueckelberg and light-cone gauge
frames can be obtained from the two-point vertex $\Gamma$ taken in the
respective Stueckelberg and light-cone gauge frames. To this end we now
discuss the spin-1 anomalous shadow field in the Stueckelberg and light-cone
gauge frames.

\subsection{ Stueckelberg gauge frame }\label{sec004-b}

For the spin-1 anomalous shadow field, the Stueckelberg gauge frame is achieved
through the use of differential constraint \rf{man24102010-22} and the
Stueckelberg gauge condition. From \rf{man24102010-24}, we see that the
scalar field $\phi_{\sh,-1}$ transforms as Stueckelberg field, i.e., this
field can be gauged away via Stueckelberg gauge fixing,
\be \label{man24102010-27a} \phi_{\sh,-1}=0\,.\ee
Using this gauge in \rf{man24102010-22}, we see that the remaining scalar
field $\phi_{\sh,1}$ can be expressed in terms of the vector field
$\phi_{\sh,0}^a$,
\be \label{man24102010-26}
\phi_{\sh,1} = -\frac{1}{r_z^{00}}\partial^a \phi_{\sh,0}^a\,.
\ee
Thus we see that the use of gauge symmetry and differential constraint
reduces field content of our approach \rf{man24102010-20} to the one in the standard approach.
In other words, the
gauge symmetry and differential constraint make it possible to match our
approach and the standard formulation of the spin-1 anomalous
shadow field.

We proceed with the discussion of Stueckelberg gauge fixed two-point vertex
of the spin-1 anomalous shadow field, i.e. we relate our vertex
\rf{man05112010-18} with the one in the standard approach to CFT. To this end
we note that vertex of the standard approach to CFT is obtained from our
gauge invariant vertex \rf{man05112010-18} by plugging Stueckelberg gauge
condition \rf{man24102010-27a} and solution to differential constraint
\rf{man24102010-26} into \rf{man05112010-19}. Doing so, we find that
two-point density $\Gamma_{12}$ \rf{man05112010-19} takes the form (up to
total derivative)
\beq
\label{man05112010-21}
&& \Gamma_{12}^{{\rm Stuck.g.fram}}  =  k_1 \Gamma_{12}^{{\rm stand}}\,,
\\[10pt]
\label{man05112010-22} && \qquad  \Gamma_{12}^{{\rm stand}}  =
\frac{\phi_\sh^a(x_1) O_{12}^{ab}\phi_\sh^b(x_2)}{|x_{12}|^{2\kappa+d}}\,,
\\[5pt]
\label{man05112010-23} && \qquad O_{12}^{ab}  \equiv \eta^{ab} -
\frac{2x_{12}^a x_{12}^b}{|x_{12}|^2}\,,
\\[5pt]
\label{man05112010-24} &&  \qquad k_1 \equiv \frac{2\kappa+d}{2(2\kappa+d-2)}\,,
\eeq
where $\Gamma_{12}^{{\rm stand}}$ \rf{man05112010-22} stands for the
two-point vertex of the spin-1 anomalous shadow field in the standard
approach to CFT. From \rf{man05112010-21}, we see that our gauge invariant
vertex taken to be in the Stueckelberg gauge frame coincides, up to
normalization factor $k_1$, with the two-point vertex in the standard
approach to CFT. As we have demonstrated in Sec.\ref{sec03S}, in the
Stueckelberg gauge frame, we are left with the vector field
$\phi_{\cur,0}^a$. Two-point correlation function of this vector field is
deined by the kernel of vertex $\Gamma^\stand$ \rf{man05112010-22}.

\subsection{ Light-cone gauge frame} \label{sec004-c}

For the spin-1 anomalous shadow field, the light-cone gauge frame is achieved
through the use of light-cone gauge and differential constraint
\rf{man24102010-22}. Taking into account gauge transformation of the
field $\phi_{\sh,0}^a$ \rf{man24102010-23}, we impose the light-cone gauge,
\be \label{man05112010-26} \phi_{\sh,0}^+ = 0\,.\ee
Using this gauge in differential constraint \rf{man24102010-22}, we obtain
solution for $\phi_\sh^-$,
\be \label{man05112010-27}  \phi_{\sh,0}^- = -
\frac{\partial^j}{\partial^+}\phi_{\sh,0}^j -
\frac{r_z^{00}}{\partial^+}\phi_{\sh,1}-
\frac{r_\zeta^{00}}{\partial^+}\Box\phi_{\sh,-1}\,. \ee
We see that we are left with vector field $\phi_{\sh,0}^i$ and the scalar
fields $\phi_{\sh,\pm1}$. These fields constitute the field content of the
light-cone gauge frame. Note that, in contrast to the Stueckelberg gauge
frame, the scalar fields $\phi_{\sh,\pm1}$ become independent field D.o.F in
the light-cone gauge frame.

Using \rf{man05112010-26} in \rf{man05112010-19} leads to light-cone
gauge fixed vertex
\beq
\label{man05112010-28}\Gamma_{12}^{({\rm l.c.})} & = &
\frac{\phi_{\sh,0}^i(x_1) \phi_{\sh,0}^i(x_2)}{2|x_{12}|^{2\kappa + d}}
\nonumber\\[5pt]
& +  & \sum_{\lambda=\pm
1}\frac{\omega_\lambda}{2|x_{12}|^{2\kappa+d-2\lambda}}
\phi_{\sh,\lambda}(x_1)\phi_{\sh,\lambda}(x_2)\,,\qquad
\eeq
where $\omega_\lambda$ are given in \rf{man05112010-19b}. As in the case of
gauge invariant vertex \rf{man05112010-19}, light-cone vertex
\rf{man05112010-28} is diagonal with respect to the fields $\phi_{\sh,0}^i$
and $\phi_{\sh,\pm1}$. Note however that, in contrast to the gauge invariant
vertex, the light-cone vertex is constructed out of the fields which are not
subject to any constraints.

Thus, as we have promised, {\it our gauge invariant vertex gives easy and
quick access to the light-cone gauge vertex}. All that is required to get
light-cone gauge vertex \rf{man05112010-28} is to replace the $so(d-1,1)$
Lorentz algebra vector indices appearing in gauge invariant vertex
\rf{man05112010-19} by the vector indices of the $so(d-2)$ algebra.

Kernel of the light-cone vertex gives two-point correlation function of
the spin-1 anomalous conformal current taken to be in the light-cone gauge.
Defining two-point correlation functions of the fields $\phi_{\cur,0}^i$,
$\phi_{\cur,\pm1}$ in a usual way,
\beq
&&  \hspace{-0.7cm} \langle \phi_{\cur,0}^i(x_1), \phi_{\cur,0}^j(x_2)\rangle =
\frac{\delta^2\Gamma^{({\rm l.c.})}}{\delta \phi_{\sh,0}^i(x_1)\delta
\phi_{\sh,0}^j(x_2)}\,,
\nonumber\\[-6pt]
&&
\\[-6pt]
&& \hspace{-0.7cm} \langle \phi_{\cur,\lambda}(x_1),
\phi_{\cur,\lambda}(x_2)\rangle = \frac{\delta^2\Gamma^{({\rm l.c.})}}{\delta
\phi_{\sh,-\lambda}(x_1)\delta \phi_{\sh,-\lambda}(x_2)}\,,\qquad
\nonumber
\eeq
$\lambda =\pm1$, and using \rf{man05112010-28}, we obtain the two-point
light-cone gauge correlation functions of the spin-1 anomalous conformal current,
\beq
&& \hspace{-0.7cm} \langle \phi_{\cur,0}^i(x_1), \phi_{\cur,0}^j(x_2)\rangle
= \frac{\delta^{ij}}{|x_{12}|^{2\kappa+d}}\,,
\nonumber\\[-5pt]
&&
\\[-5pt]
&& \hspace{-0.7cm} \langle \phi_{\cur,\lambda}(x_1),
\phi_{\cur,\lambda}(x_2)\rangle = \frac{\omega_{-\lambda}}{
|x_{12}|^{2\kappa+d+2\lambda}}\,,
\nonumber
\eeq
$\lambda=\pm 1$, where $\omega_\lambda$ are given in \rf{man05112010-19b}.

\section{ Spin-2 anomalous conformal current}
\label{sec05}

\subsection{ Gauge invariant formulation }
\label{sec05a}

To discuss gauge invariant formulation of spin-2 anomalous conformal current
in flat space of dimension $d\geq 4$ we use one rank-2 tensor field, two
vector fields, and three scalar fields,
\beq  \label{man24102010-35}
& \phi_{\cur}^{ab} &
\nonumber\\[5pt]
& \phi_{\cur,-1}^a\qquad \phi_{\cur,1}^a  &
\\[5pt]
& \phi_{\cur,-2} \qquad \phi_{\cur,0}  \qquad \phi_{\cur,2} &
\nonumber
\eeq
The fields $\phi_{\cur}^{ab}$, $\phi_{\cur,\pm1}^a$ and $\phi_{\cur,0}$,
$\phi_{\cur,\pm2}$ transform in the respective rank-2 tensor, vector and
scalar representations of the Lorentz algebra $so(d-1,1)$. Note that the
tensor field $\phi_{\cur,0}^{ab}$ is symmetric
$\phi_{\cur,0}^{ab}=\phi_{\cur,0}^{ba}$ and traceful $\phi_{\cur,0}^{aa}\ne
0$. We note that fields \rf{man24102010-35} have the conformal dimensions
\beq
\label{man24102010-36}  && \Delta_{\phi_{\cur,0}^{ab}} = \frac{d}{2} + \kappa\,,
\nonumber\\[5pt]
\label{man24102010-37}  && \Delta_{\phi_{\cur, \lambda}^a} = \frac{d}{2} +
\kappa + \lambda \,,\qquad \lambda =\pm 1\,,
\\[5pt]
\label{man24102010-38}  && \Delta_{\phi_{\cur,\lambda}} = \frac{d}{2} +
\kappa + \lambda\,,\qquad \lambda = 0,\pm 2\,,
\nonumber
\eeq
where $\kappa$ is a dimensionless parameter. In the
framework of AdS/CFT correspondence $\kappa$ is related to the mass
parameter $m$ of spin-2 massive AdS field as%
\footnote{ Parameter $\kappa$ for spin-2 field \rf{man24102010-39} should not
be confused with the one for spin-1 field \rf{man24102010-02n}.}
\be \label{man24102010-39}  \kappa = \sqrt{m^2 + \frac{d^2}{4}}\,.\ee

We now introduce the following differential constraints:
\be \label{man24102010-40}
\partial^b \phi_{\cur,0}^{ab} -\half \partial^a\phi_{\cur,0}^{bb} +
r_z^{00} \Box\phi_{\cur,-1}^a + r_\zeta^{00} \phi_{\cur,1}^a = 0 \,,\qquad
\ee
\be
\label{man24102010-41}   \partial^a \phi_{\cur,-1}^a + \half r_z^{00}
\phi_{\cur,0}^{aa} + \sqrt{2} r_z^{01} \Box \phi_{\cur,-2}
+ r_\zeta^{01} \phi_{\cur,0} = 0  \,,
\ee
\be
\label{man24102010-42}    \partial^a \phi_{\cur,1}^a + \half r_\zeta^{00} \Box
\phi_{\cur,0}^{aa} + r_z^{10}
\Box\phi_{\cur,0} + \sqrt{2} r_\zeta^{10} \phi_{\cur,2}  = 0  \,,
\ee
\beq
&& r_z^{00} \equiv
\Bigl(\frac{2\kappa+d}{4\kappa}\Bigr)^{1/2}\,,
\nonumber\\[5pt]
&& r_z^{10} \equiv
\Bigl(\frac{(2\kappa+d)(\kappa-1)d}{4 \kappa (\kappa+1)(d-2)}\Bigr)^{1/2}\,,
\nonumber\\[5pt]
&& r_z^{01} \equiv
\Bigl(\frac{2\kappa+d-2}{4(\kappa-1)}\Bigr)^{1/2}\,,
\nonumber\\[-5pt]
\label{man25102010-01} &&
\\[-5pt]
&& r_\zeta^{00} \equiv
\Bigl(\frac{2\kappa-d}{4\kappa}\Bigr)^{1/2}\,,
\nonumber\\[5pt]
&& r_\zeta^{10} \equiv
\Bigl(\frac{2\kappa-d+2}{4(\kappa+1)}\Bigr)^{1/2}\,,
\nonumber\\[5pt]
&& r_\zeta^{01} \equiv
\Bigl(\frac{(2\kappa-d)(\kappa+1)d}{4 \kappa (\kappa -
1)(d-2)}\Bigr)^{1/2}\,.
\nonumber
\eeq
One can make sure that these differential constraints are invariant under the
gauge transformations
\beq
&& \delta\phi_{\cur,0}^{ab} = \partial^a\xi_{\cur,0}^b + \partial^b
\xi_{\cur,0}^a
\nonumber\\[5pt]
&& \hspace{1cm} + \, \frac{2 r_z^{00} }{d-2}\eta^{ab} \Box \xi_{\cur,-1}
+ \frac{2 r_\zeta^{00} }{d-2}\eta^{ab} \xi_{\cur,1}\,,\qquad
\nonumber\\[5pt]
&& \delta\phi_{\cur,-1}^a =\partial^a\xi_{\cur,-1} - r_z^{00}
\xi_{\cur,0}^a\,,
\nonumber\\[5pt]
\label{man25102010-07} && \delta\phi_{\cur,1}^a =\partial^a \xi_{\cur,1} - r_\zeta^{00} \Box
\xi_{\cur,0}^a\,,
\\[5pt]
&& \delta\phi_{\cur,-2} = - \sqrt{2} r_z^{01} \xi_{\cur,-1}\,,
\nonumber\\[5pt]
&& \delta\phi_{\cur,0} = - r_\zeta^{01}\Box \xi_{\cur,-1}
- r_z^{10} \xi_{\cur,1}\,,\qquad
\nonumber\\[5pt]
&& \delta\phi_{\cur,2} = - \sqrt{2} r_\zeta^{10} \Box\xi_{\cur,1}\,,
\nonumber
\eeq
where $\xi_{\cur,0}^a$, $\xi_{\cur,\pm1}$ are gauge transformation parameters.

To complete our gauge invariant formulation we find realization of the
operator $R^a$ on space of gauge fields \rf{man24102010-35},
\beq
&& R^a\phi_{\cur,0}^{bc} =  -2\kappa r_z^{00}(\eta^{ab}\phi_{\cur,-1}^c +
\eta^{ac}\phi_{\cur,-1}^b)
\nonumber\\[5pt]
&& \hspace{1.5cm} + \,
\frac{4(\kappa-1)r_z^{00}}{d-2}\eta^{bc}\phi_{\cur,-1}^a\,,
\nonumber\\[5pt]
&& R^a\phi_{\cur,-1}^b = -2\sqrt{2}(\kappa-1)r_z^{01} \eta^{ab}
\phi_{\cur,-2}\,,
\nonumber\\[5pt]
&& R^a\phi_{\cur,1}^b = -r_\zeta^{00}(2\kappa \phi_{\cur,0}^{ab} +
\eta^{ab}\phi_{\cur,0}^{cc})
\nonumber\\[5pt]
\label{man25102010-12a1} && \hspace{1.5cm} - \, 2(\kappa+1)r_z^{10} \eta^{ab}
\phi_{\cur,0}\,,
\\[5pt]
&& R^a\phi_{\cur,-2} = 0\,,
\nonumber\\[5pt]
&& R^a\phi_{\cur,0} = -2(\kappa-1)r_\zeta^{01}\phi_{\cur,-1}^a\,,
\nonumber\\
&& R^a\phi_{\cur,2} = -2\sqrt{2}(\kappa+1)r_\zeta^{10}\phi_{\cur,1}^a\,.
\nonumber
\eeq
Using \rf{man25102010-12a1}, we check that constraints
\rf{man24102010-40}-\rf{man24102010-42} are invariant under conformal algebra
transformations \rf{04092008-01}.

\subsection{ Stueckelberg gauge frame}

For the spin-2 anomalous conformal current, the Stueckelberg gauge frame is
achieved through the use of differential constraints
\rf{man24102010-40}-\rf{man24102010-42} and Stueckelberg gauge condition.
From \rf{man25102010-07}, we see that
the vector field $\phi_{\cur,-1}^a$ and the scalar fields $\phi_{\cur,-2}$,
$\phi_{\cur,0}$ transform as Stueckelberg fields, i.e., these fields can be
gauged away via Stueckelberg gauge fixing,
\be \label{man06112010-01} \phi_{\cur,-1}^a=0\,,\qquad \phi_{\cur,-2}=0\,,
\qquad \phi_{\cur,0}=0\,.\ee
Using gauge conditions \rf{man06112010-01} in constraint \rf{man24102010-41},
we find that the field $\phi_{\cur,0}^{ab}$ becomes traceless, while using
gauge conditions \rf{man06112010-01} in constraints
\rf{man24102010-40},\rf{man24102010-42}, we find that the remaining vector
field $\phi_{\cur,1}^a$ and the scalar field $\phi_{\cur,2}$  can be
expressed in terms of the rank-2 tensor field $\phi_{\cur,0}^{ab}$,
\beq
&&  \phi_{\cur,0}^{aa}  = 0  \,,
\nonumber\\[5pt]
\label{man06112010-01a} && \phi_{\cur,1}^a =
-\frac{1}{r_\zeta^{00}}\partial^b \phi_{\cur,0}^{ab}\,,
\\[5pt]
&&  \phi_{\cur,2} = \frac{1}{\sqrt{2} r_\zeta^{00} r_\zeta^{10}}\partial^a
\partial^b \phi_{\cur,0}^{ab} \,.
\nonumber
\eeq
Relations \rf{man06112010-01}, \rf{man06112010-01a} provide the complete
description of the Stueckelberg gauge frame for the spin-2 anomalous
conformal current. We note that the traceless rank-2 tensor
$\phi_{\cur,0}^{ab}$ can be identified with the one in the standard approach
to CFT.

Thus, we see that the gauge symmetries and the differential
constraints make it possible to match our approach and the standard one,
i.e., by  gauging away the Stueckelberg fields \rf{man06112010-01} and by
solving differential constraints \rf{man24102010-40}-\rf{man24102010-42} we
obtain the standard formulation of the spin-2 anomalous conformal current.

\subsection{ Light-cone gauge frame }

For the spin-2 anomalous conformal current, the light-cone gauge frame is
achieved through the use of differential constraints
\rf{man24102010-40}-\rf{man24102010-42} and light-cone gauge condition.

Using the gauge transformations of the fields $\phi_{\cur,0}^{ab}$,
$\phi_{\cur,\pm 1}^a$ \rf{man25102010-07}, we impose the
light-cone gauge,
\be \phi_{\cur,0}^{+a} =0 \,,\qquad \phi_{\cur,\lambda}^+ =0\,,\qquad \lambda
= \pm 1 \,.\ee
Plugging this gauge in differential constraints
\rf{man24102010-40}-\rf{man24102010-42}, we find
\beq
&& \hspace{-0.3cm}\phi_{\cur,0}^{ii} = 0 \,,
\nonumber\\[5pt]
&&  \hspace{-0.3cm} \phi_{\cur,0}^{-i} = -
\frac{\partial^j}{\partial^+}\phi_{\cur,0}^{ij}
-\frac{r_z^{00}}{\partial^+}\Box \phi_{\cur,-1}^i -
\frac{r_\zeta^{00}}{\partial^+}\phi_{\cur,1}^i \,,
\nonumber\\
\nonumber\\[3pt]
&& \hspace{-0.3cm} \phi_{\cur,0}^{--} =
\frac{\partial^i\partial^j}{\partial^+\partial^+}\phi_{\cur,0}^{ij}
+ \frac{2r_z^{00}\partial^i}{\partial^+\partial^+}\Box\phi_{\cur,-1}^i
+ \frac{2r_\zeta^{00}\partial^i}{\partial^+\partial^+}\phi_{\cur,1}^i
\nonumber\\[5pt]
&& \hspace{0.7cm}  + \,
\frac{\sqrt{2}r_z^{00}r_z^{10}}{\partial^+\partial^+}\Box^2\phi_{\cur,-2} +
\frac{\sqrt{2}r_\zeta^{00}r_\zeta^{01}}{\partial^+\partial^+}\phi_{\cur,2}
\nonumber\\[5pt]
&& \hspace{0.7cm} + \,
\frac{r_z^{00}r_\zeta^{01}+r_\zeta^{00}r_z^{10}}{\partial^+\partial^+}\Box\phi_{\cur,0}\,,
\nonumber\\[5pt]
&&  \hspace{-0.3cm} \phi_{\cur,-1}^- = -
\frac{\partial^j}{\partial^+}\phi_{\cur,-1}^j - \frac{\sqrt{2}
r_z^{01}}{\partial^+}\Box \phi_{\cur,-2} -
\frac{r_\zeta^{01}}{\partial^+}\phi_{\cur,0}\,,
\nonumber\\
\nonumber\\
&& \hspace{-0.3cm} \phi_{\cur,1}^-  = -
\frac{\partial^j}{\partial^+}\phi_{\cur,1}^j -  \frac{
r_z^{10}}{\partial^+}\Box \phi_{\cur,0} - \frac{\sqrt{2}
r_\zeta^{10}}{\partial^+}\phi_{\cur,2}\,.
\nonumber\\
\eeq
We see that we are left with $so(d-2)$ algebra traceless rank-2 tensor field,
two vector fields, and three scalar fields,
\beq
& \phi_{\cur,0}^{ij}  &
\nonumber\\[5pt]
\label{man07112010-16} & \phi_{\cur,-1}^i   \qquad \phi_{\cur,1}^i  &
\\[5pt]
& \phi_{\cur,-2} \qquad \phi_{\cur,0}  \qquad \phi_{\cur,2} &
\nonumber
\eeq
which constitute field content of the light-cone
gauge frame.

\section{ Spin-2 anomalous shadow field}
\label{sec06}

\subsection{ Gauge invariant formulation }

To discuss gauge invariant formulation of spin-2 anomalous shadow field in
flat space of dimension $d\geq 4$ we use one rank-2 tensor field, two vector fields, and
three scalars fields,
\beq
& \phi_{\sh,0}^{ab}  &
\nonumber\\[5pt]
\label{man25102010-14} & \phi_{\sh,-1}^a \qquad \phi_{\sh,1}^a  &
\\[5pt]
& \phi_{\sh,-2} \qquad \phi_{\sh,0}  \qquad \phi_{\sh,2} &
\nonumber
\eeq
The fields $\phi_{\sh,0}^{ab}$, $\phi_{\sh,\pm 1}^a$ and $\phi_{\sh,0}$,
$\phi_{\sh,\pm2}$ transform in the respective rank-2 tensor, vector and
scalar representations of the Lorentz algebra $so(d-1,1)$.  Note that the
tensor field $\phi_{\sh,0}^{ab}$ is symmetric
$\phi_{\sh,0}^{ab}=\phi_{\sh,0}^{ba}$ and traceful $\phi_{\sh,0}^{aa}\ne 0$.
Conformal dimensions of the fields are given by
\beq
&& \Delta_{\phi_{\sh,0}^{ab}} = \frac{d}{2} - \kappa\,,
\nonumber\\[5pt]
\label{man25102010-14n} && \Delta_{\phi_{\sh, \lambda}^a} = \frac{d}{2} -
\kappa + \lambda \,,\qquad \lambda =\pm 1\,,
\\[5pt]
&& \Delta_{\phi_{\sh,\lambda}} = \frac{d}{2} - \kappa + \lambda\,,\qquad
\lambda = 0,\pm 2\,,
\nonumber
\eeq
In the framework of AdS/CFT correspondence, $\kappa$ is related to the mass
parameter $m$ of spin-2 massive AdS field as in \rf{man24102010-39}.

We now introduce the following differential constraints:
\beq \label{man25102010-15}
&& \partial^b \phi_{\sh,0}^{ab} -\half \partial^a\phi_{\sh,0}^{bb} +
r_\zeta^{00} \Box \phi_{\sh,-1}^a + r_z^{00} \phi_{\sh,1}^a  = 0 \,,
\nonumber\\
\\
&& \label{man25102010-16} \partial^a \phi_{\sh,-1}^a + \half r_\zeta^{00}
\phi_{\sh,0}^{aa} + \sqrt{2} r_\zeta^{10} \Box \phi_{\sh,-2} + r_z^{10}
\phi_{\sh,0} = 0,
\nonumber\\
\\
&& \label{man25102010-17} \partial^a \phi_{\sh,1}^a + \half r_z^{00} \Box
\phi_{\sh,0}^{aa} + r_\zeta^{01} \Box
\phi_{\sh,0} + \sqrt{2} r_z^{01} \phi_{\sh,2}  = 0  \,,
\nonumber\\
\eeq

\vspace{-0.2cm}
\noindent where the parameters $r_\zeta^{mn}$ and $r_z^{mn}$ are given in
\rf{man25102010-01}. One can make sure that these
constraints are invariant under the gauge transformations
\beq
&& \delta\phi_{\sh,0}^{ab} = \partial^a\xi_{\sh,0}^b + \partial^b
\xi_{\sh,0}^a
\nonumber\\[5pt]
&& \hspace{1cm} + \, \frac{2 r_z^{00} }{d-2}\eta^{ab} \xi_{\sh,1} + \frac{2
r_\zeta^{00}}{d-2}\eta^{ab} \Box\xi_{\sh,-1}\,,
\nonumber\\[5pt]
&& \delta\phi_{\sh,-1}^a = \partial^a \xi_{\sh,-1} - r_\zeta^{00}
\xi_{\sh,0}^a\,,
\nonumber\\[5pt]
\label{man29102010-03} && \delta\phi_{\sh,1}^a =\partial^a\xi_{\sh,1} - r_z^{00} \Box
\xi_{\sh,0}^a\,,
\\[5pt]
&& \delta\phi_{\sh,-2} = - \sqrt{2} r_\zeta^{10} \xi_{\sh,-1}\,,
\nonumber\\[5pt]
&& \delta\phi_{\sh,0} = - r_\zeta^{01} \xi_{\sh,1} - r_z^{10} \Box \xi_{\sh,-1}\,,\qquad
\nonumber\\[5pt]
&&
\delta\phi_{\sh,2} = - \sqrt{2} r_z^{01} \Box \xi_{\sh,1}\,,
\nonumber
\eeq
where $\xi_{\sh,0}^a$, $\xi_{\sh,\pm 1}$ are gauge transformation parameters.

We then find that a realization of the
operator $R^a$ on fields \rf{man25102010-14} takes the following form:
\beq
&& R^a\phi_{\sh,0}^{bc} =  2\kappa r_\zeta^{00}(\eta^{ab}\phi_{\sh,-1}^c +
\eta^{ac}\phi_{\sh,-1}^b)
\nonumber\\[5pt]
&& \hspace{1.5cm} - \,
\frac{4(\kappa+1)r_\zeta^{00}}{d-2}\eta^{bc}\phi_{\sh,-1}^a\,,
\nonumber\\[5pt]
 && R^a\phi_{\sh,-1}^b =
2\sqrt{2}(\kappa+1)r_\zeta^{10} \eta^{ab} \phi_{\sh,-2}\,,
\nonumber\\[5pt]
\label{man11112010-04} && R^a\phi_{\sh,1}^b = r_z^{00}(2\kappa
\phi_{\sh,0}^{ab} - \eta^{ab}\phi_{\sh,0}^{cc})
\\[5pt]
&& \hspace{1.5cm} + \,  2(\kappa-1)r_\zeta^{01} \eta^{ab} \phi_{\sh,0}\,,
\nonumber\\[5pt]
&& R^a\phi_{\sh,-2} = 0\,,
\nonumber\\[5pt]
&& R^a\phi_{\sh,0} =
2(\kappa+1)r_z^{10}\phi_{\sh,-1}^a\,,
\nonumber\\[5pt]
&& R^a\phi_{\sh,2} =
2\sqrt{2}(\kappa-1)r_z^{01}\phi_{\sh,1}^a\,.
\nonumber
\eeq
Using \rf{man11112010-04}, we check that constraints
\rf{man25102010-15}-\rf{man25102010-17} are invariant under transformations
of the conformal algebra.

We proceed with the discussion of two-point vertex for the spin-2 anomalous
shadow field. The gauge invariant two-point vertex we find takes the form
given \rf{man05112010-18}, where the two-point density $\Gamma_{12}$ is given
by
\beq \label{man07112010-04}
\Gamma_{12} &  = &
\frac{1}{4|x_{12}|^{2\kappa+d}}\Bigl(\phi_{\sh,0}^{ab}(x_1)\phi_{\sh,0}^{ab}(x_2)
\nonumber\\
&& \hspace{2.3cm} - \half
\phi_{\sh,0}^{aa}(x_1)\phi_{\sh,0}^{bb}(x_2)\Bigr)
\nonumber\\[5pt]
& +  & \sum_{\lambda=\pm 1}\frac{\omega_\lambda}{2|x_{12}|^{2\kappa + d
-2\lambda}} \phi_{\sh,\lambda}^a(x_1)\phi_{\sh,\lambda}^a(x_2)
\nonumber\\[5pt]
& +  & \sum_{\lambda=0,\pm 2}\frac{\omega_\lambda}{2|x_{12}|^{2\kappa + d
-2\lambda}} \phi_{\sh,\lambda}(x_1)\phi_{\sh,\lambda}(x_2)\,,\qquad
\\[3pt]
&& \hspace{-0.7cm}  \omega_1 = \frac{1}{2\kappa(2\kappa+d-2)}\,,\qquad\quad
\omega_0 = 1\,,
\nonumber\\[3pt]
&& \hspace{-0.7cm}  \omega_{-1} = 2(\kappa+1)(2\kappa+d)\,,
\nonumber\\[-5pt]
\label{man07112010-04a1} &&
\\[-5pt]
&& \hspace{-0.7cm}  \omega_2 =
\frac{1}{4\kappa(\kappa-1)(2\kappa+d-2)(2\kappa+d-4)}\,,
\nonumber\\[5pt]
&& \hspace{-0.7cm} \omega_{-2} =
4(\kappa+1)(\kappa+2)(2\kappa+d)(2\kappa+d+2)\,.\qquad\quad
\nonumber
\eeq
We check that this vertex is invariant under both gauge transformations
\rf{man29102010-03} and global conformal transformations of the spin-2 anomalous
shadow field. Remarkable feature of the vertex is its diagonal form with
respect to the gauge fields entering field content \rf{man25102010-14}.

\subsection{ Stueckelberg gauge frame }\label{sec006-b}

For the spin-2 anomalous shadow field, the Stueckelberg gauge frame is
achieved though the use of differential constraints
\rf{man25102010-15}-\rf{man25102010-17} and Stueckelberg gauge condition.
From gauge transformations \rf{man29102010-03}, we see that the vector field
$\phi_{\sh,-1}^a$ and the scalar fields $\phi_{\sh,-2}$, $\phi_{\sh,0}$
transform as Stueckelberg fields, i.e., these fields can be gauged away via
Stueckelberg gauge fixing,
\be \label{man06112010-02}
\phi_{\sh,-1}^a=0\,,\qquad \phi_{\sh,-2}=0\,, \qquad \phi_{\sh,0}=0\,.\ee
Using gauge conditions \rf{man06112010-02} in constraint \rf{man25102010-16},
we find that the field $\phi_{\sh,0}^{ab}$ becomes traceless, while using
gauge conditions \rf{man06112010-02} in constraints
\rf{man25102010-15},\rf{man25102010-17} we find that the remaining vector
field $\phi_{\sh,1}^a$ and the scalar field $\phi_{\sh,2}$  can be expressed
in terms of the rank-2 tensor field $\phi_{\sh,0}^{ab}$,
\beq
&& \phi_{\sh,0}^{aa}  = 0,
\nonumber\\[3pt]
\label{man07112010-03} && \phi_{\sh,1}^a = -\frac{1}{r_z^{00}}\partial^b
\phi_{\sh,0}^{ab}\,,
\\[3pt]
&& \phi_{\sh,2} = \frac{1}{\sqrt{2} r_z^{00} r_z^{01}}\partial^a\partial^b
\phi_{\sh,0}^{ab} \,.
\nonumber
\eeq
Relations \rf{man06112010-02}, \rf{man07112010-03} provide the complete
description of the Stueckelberg gauge frame for the spin-2 anomalous shadow
field.

Plugging \rf{man06112010-02}, \rf{man07112010-03} in \rf{man07112010-04}, we
find that our $\Gamma_{12}$ \rf{man07112010-04} takes  the form (up to toal
derivative),
\beq \label{man07112010-05}
&& \Gamma_{12}^{{\rm Stuck.g.fram}}  =   k_2 \Gamma_{12}^{{\rm stand}}\,,
\\[5pt]
\label{man07112010-06} && \qquad \Gamma_{12}^{{\rm stand}}  =
\phi_{\sh,0}^{a_1a_2}(x_1) \frac{ O_{12}^{a_1b_1}
O_{12}^{a_2b_2}}{|x_{12}|^{2\kappa+d}} \phi_{\sh,0}^{b_1b_2}(x_2) \,,\qquad
\quad
\\[5pt]
\label{man07112010-07} && \qquad k_2 \equiv
\frac{2\kappa+d+2}{4(2\kappa+d-2)}\,,
\eeq
where $O_{12}^{ab}$ is defined in \rf{man05112010-23}, while
$\Gamma_{12}^{{\rm stand}}$ \rf{man07112010-06} stands for the two-point
vertex of the spin-2 anomalous shadow field in the standard approach to CFT.
From \rf{man07112010-05}, we see that our gauge invariant vertex taken to be
in the Stueckelberg gauge frame coincides, up to normalization factor $k_2$,
with the two-point vertex in the standard approach to CFT. Kernel of vertex
$\Gamma^{{\rm stand}}$ \rf{man07112010-06} defines two-point correlation
function of the spin-2 conformal conformal current taken to be in the
Stueckelberg gauge frame.

\subsection{ Light-cone gauge frame}\label{sec006-c}

For the spin-2 anomalous shadow field, the light-cone gauge frame is achieved
through the use of differential constraints
\rf{man25102010-15}-\rf{man25102010-17} and light-cone gauge. Taking into
account the gauge transformations of the fields $\phi_{\sh,0}^{ab}$,
$\phi_{\sh,\pm1}^a$ given in \rf{man29102010-03}, we
impose the light-cone gauge condition,
\be \label{man07112010-08} \phi_{\sh,0}^{+a} = 0\,, \qquad
\phi_{\sh,\lambda}^+ = 0\,,\qquad \lambda =\pm 1\,. \ee
Plugging this gauge condition in constraints
\rf{man25102010-15}-\rf{man25102010-17}, we find
\beq
&& \hspace{-1.1cm} \phi_{\sh,0}^{ii} = 0 \,,
\nonumber\\[5pt]
&& \hspace{-1.1cm} \phi_{\sh,0}^{-i} = -
\frac{\partial^j}{\partial^+}\phi_{\sh,0}^{ij}
-\frac{r_\zeta^{00}}{\partial^+}\Box \phi_{\sh,-1}^i -
\frac{r_z^{00}}{\partial^+} \phi_{\sh,1}^i \,,
\nonumber\\[5pt]
&& \hspace{-1.1cm} \phi_{\sh,0}^{--} =
\frac{\partial^i\partial^j}{\partial^+\partial^+}\phi_{\sh,0}^{ij}
+ \frac{2r_\zeta^{00}\partial^i}{\partial^+\partial^+}\Box\phi_{\sh,-1}^i
+ \frac{2r_z^{00}\partial^i}{\partial^+\partial^+}\phi_{\sh,1}^i
\nonumber\\[5pt]
&&  \hspace{-0.1cm} + \,
\frac{\sqrt{2}r_\zeta^{00}r_\zeta^{10}}{\partial^+\partial^+}\Box^2\phi_{\sh,-2}
+ \frac{\sqrt{2}r_z^{00}r_z^{01}}{\partial^+\partial^+}\phi_{\sh,2}
\nonumber\\[5pt]
&&  \hspace{-0.1cm} + \,
\frac{r_z^{00}r_\zeta^{01}+r_\zeta^{00}r_z^{10}}{\partial^+\partial^+}\Box\phi_{\sh,0}\,,
\nonumber\\[5pt]
&& \hspace{-1.1cm} \phi_{\sh,-1}^- = -
\frac{\partial^j}{\partial^+}\phi_{\sh,-1}^j - \frac{\sqrt{2}
r_\zeta^{10}}{\partial^+}\Box \phi_{\sh,-2} -
\frac{r_z^{10}}{\partial^+}\phi_{\sh,0}\,,
\nonumber\\[5pt]
\label{man07112010-09} && \hspace{-1.1cm} \phi_{\sh,1}^-  = -
\frac{\partial^j}{\partial^+}\phi_{\sh,1}^j - \frac{
r_\zeta^{01}}{\partial^+}\Box \phi_{\sh,0} - \frac{\sqrt{2}
r_z^{01}}{\partial^+}\phi_{\sh,2}\,.
\eeq
We see that we are left with the $so(d-2)$ algebra traceless rank-2 tensor
field, two vector fields and three scalar fields,
\beq
& \phi_{\sh,0}^{ij}  &
\nonumber\\[5pt]
\label{man07112010-14} & \phi_{\sh,-1}^i\qquad \phi_{\sh,1}^i &
\\[5pt]
& \phi_{\sh,-2} \qquad \phi_{\sh,0}  \qquad \phi_{\sh,2} &
\nonumber
\eeq
which constitute a field content of the spin-2 anomalous shadow field in
light-cone gauge frame. Note that, in contrast to the Stueckelberg gauge
frame, the vector fields and the scalar fields become independent field D.o.F
in the light-cone gauge frame.

Using \rf{man07112010-08} in \rf{man07112010-04},
leads to light-cone gauge fixed vertex
\beq \label{man07112010-15}
\Gamma_{12}^{({\rm l.c.})} &  = &
\frac{1}{4|x_{12}|^{2\kappa+d}}\phi_{\sh,0}^{ij}(x_1)\phi_{\sh,0}^{ij}(x_2)
\nonumber\\[5pt]
& +  & \sum_{\lambda=\pm 1}\frac{\omega_\lambda}{2|x_{12}|^{2\kappa + d -
2\lambda}}  \phi_{\sh,\lambda}^i(x_1)\phi_{\sh,\lambda}^i(x_2)
\nonumber\\[5pt]
& +  & \sum_{\lambda=0,\pm 2}\frac{\omega_\lambda}{2|x_{12}|^{2\kappa + d -2
\lambda}} \phi_{\sh,\lambda}(x_1)\phi_{\sh,\lambda}(x_2)\,, \qquad\quad
\eeq
where $\omega_\lambda$ are defined in \rf{man07112010-04a1}. We see that, as
in the case of gauge invariant vertex \rf{man07112010-04}, light-cone vertex
\rf{man07112010-15} is diagonal with respect to the fields entering the field
content of light-cone gauge frame \rf{man07112010-14}. Note however that, in
contrast to the gauge invariant vertex, the light-cone vertex is constructed
out of the fields \rf{man07112010-14} which are not subject to any
differential constraints.

As before, we see that {\it our gauge invariant vertex gives easy and quick
access to the light-cone gauge vertex}. Namely, all that is required to get light-cone
gauge vertex \rf{man07112010-15} is to remove trace of the tensor field
$\phi_{\sh,0}^{ab}$ and replace the $so(d-1,1)$ Lorentz algebra vector
indices appearing in gauge invariant vertex \rf{man07112010-04} by the vector
indices of the $so(d-2)$ algebra.

Kernel of light-cone vertex \rf{man07112010-15} gives two-point correlation
function of the spin-2 anomalous conformal current taken to be in the
light-cone gauge. Defining two-point correlation functions for light-cone
fields of the anomalous conformal current \rf{man07112010-16} in usual way
\beq
&&  \hspace{-0.7cm} \langle \phi_{\cur,0}^{ij}(x_1), \phi_{\cur,0}^{kl}j(x_2)\rangle \equiv
\frac{\delta^2\Gamma^{({\rm l.c.})}}{\delta \phi_{\sh,0}^{ij}(x_1)\delta
\phi_{\sh,0}^{kl}(x_2)}\,,
\nonumber\\[5pt]
&& \hspace{-0.7cm} \langle \phi_{\cur,\lambda}^i(x_1), \phi_{\cur,\lambda}^j(x_2)\rangle \equiv
\frac{\delta^2\Gamma^{({\rm l.c.})}}{\delta \phi_{\sh,-\lambda}^i(x_1)\delta
\phi_{\sh,-\lambda}^j(x_2)}\,,\qquad
\\[5pt]
&& \hspace{-0.7cm} \langle \phi_{\cur,\lambda}(x_1), \phi_{\cur,\lambda}(x_2)\rangle \equiv
\frac{\delta^2\Gamma^{({\rm l.c.})}}{\delta \phi_{\sh,-\lambda}(x_1)\delta
\phi_{\sh,-\lambda}(x_2)}\,,\qquad
\nonumber
\eeq
we obtain
\beq
&&  \hspace{-0.8cm} \langle \phi_{\cur,0}^{ij}(x_1),
\phi_{\cur,0}^{kl}(x_2)\rangle = \frac{1}{|x_{12}|^{2\kappa+d} }
\Pi^{ij;kl}\,,
\nonumber\\[5pt]
&& \hspace{-0.8cm} \langle \phi_{\cur,\lambda}^i(x_1),
\phi_{\cur,\lambda}^j(x_2)\rangle =
\frac{\omega_{-\lambda}}{|x_{12}|^{2\kappa+d+ 2\lambda} } \delta^{ij}\,,
\\[5pt]
&& \hspace{-0.8cm} \langle \phi_{\cur,\lambda}(x_1),
\phi_{\cur,\lambda}(x_2)\rangle =
\frac{\omega_{-\lambda}}{|x_{12}|^{2\kappa+d+2\lambda}}\,, \qquad
\nonumber
\eeq
where $\omega_\lambda$ are defined in \rf{man07112010-04a1} and we use the
notation
\be
\qquad \Pi^{ij;kl} = \half\Bigl(\delta^{ik}\delta^{jl} +
\delta^{il}\delta^{jk} - \frac{2}{d-2} \delta^{ij}\delta^{kl}\Bigr) \,. \ee

\section{ Two point current-shadow field interaction vertex}
\label{sec09}

We now discuss two-point current-shadow field interaction vertex. In the
gauge invariant approach, the interaction vertex is determined by requiring
the vertex to be invariant under both gauge transformations of currents and
shadow fields. Also, the interaction vertex should be invariant under
conformal algebra transformations.

{\bf Spin-1}. We begin with spin-1 fields. Let us consider the following
vertex:
\be  \label{man29102010-07}   \LL = \phi_{\cur,0}^a \phi_{\sh,0}^a +
\phi_{\cur,-1}\phi_{\sh,1}+
\phi_{\cur,1}\phi_{\sh,-1}\,. \ee
Denoting the left hand side of \rf{man24102010-22} by $C_{\sh}$ we find that
under gauge transformations of the current
\rf{man24102010-06}-\rf{man24102010-08} the variation of vertex
\rf{man29102010-07} takes the form (up to total derivative)
\be  \label{man29102010-08} \delta_{ \xi_{\cur,0}} \LL = - \xi_{\cur,0} C_\sh\,. \ee
From this expression, we see that the vertex $\LL$ is invariant under gauge
transformations of the current provided the shadow field satisfies
differential constraint \rf{man24102010-22}. Denoting the left hand side of
\rf{man24102010-03} by $C_{\cur}$ we find that under gauge transformations of
the shadow field  \rf{man24102010-23}-\rf{man24102010-25} the variation of
vertex \rf{man29102010-07} takes the form (up to total derivative)
\be \label{man29102010-09} \delta_{ \xi_{sh}} \LL =  -\xi_{\sh,0} C_\cur \,, \ee
i.e., the vertex $\LL$ is invariant under gauge transformations of the shadow
field provided the current satisfies differential constraint
\rf{man24102010-03}.

Making use of the realization of the conformal algebra symmetries
obtained in the Sections \ref{sec03},\ref{sec04} we check that vertex $\LL$
\rf{man29102010-07} is invariant under the conformal algebra
transformations.

{\bf Spin-2}. We proceed with spin-2 fields. One can make sure that the
following vertex
\beq
\label{man29102010-10}
&&
\LL = \half \phi_{\cur,0}^{ab}\phi_{\sh,0}^{ab} -
\frac{1}{4} \phi_{\cur,0}^{aa}\phi_{\sh,0}^{bb}
\nonumber\\[5pt]
&& + \sum_{\lambda = \pm 1}\phi_{\cur,\lambda}^a \phi_{\sh,-\lambda}^a +
\sum_{\lambda = 0,\pm 2} \phi_{\cur,\lambda} \phi_{\sh,-\lambda} \qquad \eeq
is invariant under gauge transformations of the spin-2 shadow field
\rf{man29102010-03} provided the spin-2 current satisfies
differential constraints \rf{man24102010-40}-\rf{man24102010-42}. Vertex
\rf{man29102010-10} is also invariant under gauge transformations of the spin-2
anomalous current \rf{man25102010-07} provided the spin-2 shadow field
satisfies differential constraints \rf{man25102010-15}-\rf{man25102010-17}.
Using the representation for generators of the conformal algebra obtained in
the Sections \ref{sec05},\ref{sec06} we check that vertex $\LL$
\rf{man29102010-10} is invariant under the conformal algebra transformations.

\section{AdS/CFT correspondence. Preliminaries}\label{secAdS/CFT}

We now study AdS/CFT correspondence for free
massive AdS fields and boundary anomalous conformal currents and shadow
fields. To this end we use the gauge invariant CFT adapted description of
AdS massive fields and modified Lorentz and de Donder gauges found in
Ref.\cite{Metsaev:2009hp}. {\it It is the use of our fields and the modified
Lorentz and de Donder gauges that leads to decoupled form of gauge fixed
equations of motion and surprisingly simple Lagrangian}%
\footnote{Our massive gauge fields are obtained from gauge fields used in the
standard gauge invariant approach to massive fields by the invertible
transformation. Details of the transformation may be found in Appendices
\ref{app-01},\ref{app-02}. Discussion of interesting methods for solving
AdS field equations of motion without gauge fixing may be found in
Refs.\cite{Bolotin:1999fa,Didenko:2009td}.}.
Owing these properties of our fields and the modified (Lorentz) de Donder
gauge, we
simplify significantly the computation of the effective action%
\footnote{ We remind that the bulk action evaluated on solution of the
Dirichlet problem is referred to as effective action in this paper.}. Note that the
modified (Lorentz) de Donder gauge turns out to be invariant under on-shell leftover
gauge symmetries of bulk AdS fields. Also note that, in our approach, we have
gauge symmetries not only at AdS side but also at the boundary CFT. Therefore,
in the framework of our approach, the study of AdS/CFT correspondence implies
matching of:
\\
{\bf i}) Lorentz (de Donder) gauge conditions for bulk massive fields and
differential constraints for boundary anomalous conformal currents and shadow
fields;
\\
{\bf ii}) leftover on-shell gauge symmetries for bulk massive fields and
gauge symmetries of boundary anomalous conformal currents and shadow fields;
\\
{\bf iii}) on-shell global symmetries of bulk massive fields and global
symmetries of boundary anomalous conformal currents and shadow fields;
\\
{\bf iv}) effective action evaluated on solution of equations of motion with
the Dirichlet problem corresponding to the boundary anomalous shadow field
and boundary two-point gauge invariant vertex for anomalous shadow field.

{\bf Global AdS symmetries in CFT adapted approach}. Relativistic symmetries
of the $AdS_{d+1}$ field dynamics are described by the $so(d,2)$ algebra. In
$d$-dimensional space, global symmetries of anomalous conformal currents and
shadow fields are also described by the $so(d,2)$ algebra. To discuss global
symmetries of anomalous conformal currents and shadow fields we have used
conformal basis of the $so(d,2)$ algebra (see \rf{man12112010-01}). Therefore
for application to the study of AdS/CFT correspondence, it is convenient to
realize the relativistic bulk $so(d,2)$ algebra symmetries by using basis of
the conformal algebra. Most convenient way to achieve conformal basis
realization of bulk $so(d,2)$ symmetries is to use Poincar\'e parametrization
of AdS
space%
\footnote{ In our approach only $so(d-1,1)$ symmetries are realized
manifestly. The $so(d,2)$ symmetries could be realized manifestly by using
ambient space approach (see e.g.
\cite{Metsaev:1995re,Bekaert:2009fg,Bonezzi:2010jr})},
\be \label{lineelem01} ds^2 = \frac{1}{z^2}(dx^a dx^a + dz\, dz)\,. \ee
In this parametrization, the $so(d,2)$ algebra transformations of the massive
arbitrary spin AdS field $\phi$ take the form $\delta_{\hat{G}}\phi = \hat{G}
\phi$, where realization of the $so(d,2)$ algebra generators $\hat{G}$ in
terms of differential operators acting on $\phi$ is given by
\beq
\label{conalggenlis01ads} && P^a = \partial^a \,,
\\[3pt]
\label{conalggenlis02ads} && J^{ab} = x^a\partial^b -  x^b\partial^a +
M^{ab}\,,
\\[3pt]
\label{conalggenlis03ads} && D = x\partial + \Delta\,, \qquad \Delta =
z\partial_z + \frac{d-1}{2}\,, \qquad\quad
\\[3pt]
\label{conalggenlis04ads} && K^a = K_{\Delta,M}^a + R^a\,,
\\[3pt]
&& \hspace{1cm} K_{\Delta,M}^a= -\half x^2\partial^a + x^a D + M^{ab}x^b
\,,\qquad
\\[3pt]
\label{14092008-06} &&  \hspace{1cm} R^a  = R_\smzero^a + R_\smone^a\,,
\\[3pt]
\label{14092008-08} && \hspace{1cm} R_\smone^a =  -\half z^2 \partial^a\,.
\eeq
Operator $R_\smzero^a$ \rf{14092008-06} does not depend on boundary
coordinates $x^a$, boundary derivatives $\partial^a$, and derivative with
respect to radial coordinate, $\partial_z$. Operator $R_\smzero^a$ acting on
spin D.o.F. depends only on the radial coordinate $z$. Thus, we see all that
is required to complete description of the global symmetries of AdS field
dynamics is to find realization of the operator $R_\smzero^a$ on space of
gauge AdS fields.

{\bf AdS/CFT correspondence for spin-0 anomalous current and normalizable
modes of scalar massive AdS field}%
\footnote{Also see Refs.\cite{Balasubramanian:1998sn}.}.
Because use  of modified Lorentz (de Donder) gauge makes study of AdS/CFT
correspondence for spin-1 (spin-2) field similar to the one for scalar field
we begin with brief review of the AdS/CFT correspondence for the scalar
field.

Action and Lagrangian for the
massive scalar field in $AdS_{d+1}$ background take the form%
\footnote{ From now on we use, unless otherwise specified, the Euclidian
signature.}
\beq \label{19072009-01}
& S =  \int d^dx dz\,  \LL \,, &
\\[7pt]
\label{19072009-02} & \LL = \half \sqrt{|g|}\left(g^{\mu\nu}\partial_\mu \Phi
\partial_\nu \Phi + m^2 \Phi^2 \right)\,. & \eeq
In terms of the canonical normalized field $\phi$ defined by relation $\Phi=
z^{\frac{d-1}{2}}\phi$, the Lagrangian takes the form (up to total
derivative)
\beq
\label{19072009-04}
&& \LL =  \half |d\phi|^2 + \half |\TT_{\nu -\half } \phi|^2\,,
\\[5pt]
\label{09072009-04} && \hspace{1cm} \TT_\nu \equiv \partial_z +
\frac{\nu}{z}\,,
\\[5pt]
\label{19072009-05} && \hspace{1cm} \nu = \sqrt{m^2 + \frac{d^2}{4}}\,.
\eeq
Equation of motion obtained from Lagrangian \rf{19072009-04} takes the form
\beq \label{19072009-06}
&& \Box_\nu  \phi = 0 \,,
\\[5pt]
&& \label{09072009-09}
\Box_\nu \equiv \Box + \partial_z^2 - \frac{1}{z^2}(\nu^2 -\frac{1}{4})\,.
\eeq
Normalizable solution of equation \rf{19072009-06} is given by
\beq
\label{man16112010-01} && \phi(x,z) = U_\nu^\sc \phi_\cur(x)\,,
\\[5pt]
\label{man01112010-31x} && \hspace{1cm} U_\nu^\sc \equiv h_\nu \sqrt{zq}
J_\nu(zq) q^{-(\nu + \half)}\,,
\\[5pt]
&& \hspace{1cm} h_\nu\equiv 2^\nu\Gamma(\nu+1)\,,\qquad  q^2\equiv
\Box\,,\qquad \eeq
where $J_\nu$ stands for the Bessel function. The asymptotic behavior of
solution \rf{man16112010-01} is given by
\be \label{man16112010-02}
\phi(x,z) \ \ \stackrel{z\rightarrow 0}{\longrightarrow} \ \ z^{\nu + \half}
\phi_\cur(x)\,, \ee
i.e., we see that spin-0 current $\phi_\cur$ is indeed boundary value of the
normalizable solution.

In the case under consideration, we have no gauge symmetries and gauge
conditions. Therefore all that is required to complete AdS/CFT correspondence
is to match bulk global symmetries of the AdS field $\phi(x,z)$ and boundary
global symmetries of the current $\phi_\cur(x)$. Global symmetries on AdS
side and CFT side are described in \rf{conalggenlis01ads}-\rf{14092008-08}
and \rf{conalggenlis01}-\rf{conalggenlis04} respectively. We see that the
Poincar\'e symmetries match automatically. Using the notation $D_{_{AdS}}$
and $D_{_{CFT}}$ to indicate the respective realizations of $D$ symmetry on
bulk fields \rf{conalggenlis03ads} and conformal currents \rf{conalggenlis03}
we obtain the relation
\be
D_{_{AdS}} \phi(x,z) =  U_\nu^\sc D_{_{CFT}} \phi_\cur(x)\,,
\ee
where the expressions for $D_{_{CFT}}$ corresponding to $\phi_\cur$ can be
obtained from \rf{conalggenlis03} by using $\Delta = \frac{d}{2}+\nu$ with
$\nu$ given in \rf{19072009-05}. Thus, $D$ symmetries of $\phi(x,z)$ and
$\phi_\cur(x)$ also match. To match the $K^a$ symmetries in
\rf{conalggenlis04} and \rf{conalggenlis04ads} we note that the respective
operators $R_\smzero^a$ and $R^a$ act trivially, $R_\smzero^a\phi(x,z)=0$,
$R^a\phi_\cur(x)=0$ and then make sure that the $K^a$ symmetries also match.

{\bf AdS/CFT correspondence for spin-0 shadow field and non-normalizable
modes of scalar massive AdS field}. Following the procedure in
Ref.\cite{wit}, we note that non-normalizable solution of equation
\rf{19072009-06} with the Dirichlet problem corresponding to boundary shadow
scalar field $\phi_\sh(x)$ takes the form
\beq \label{19072009-09}
\phi(x,z) & = & \sigma \int d^dy\, G_\nu (x-y,z) \phi_\sh(y)\,,
\\[3pt]
\label{10072009-12} && G_\nu(x,z) = \frac{c_\nu z^{\nu+\half}}{ (z^2+
|x|^2)^{\nu + \frac{d}{2}} }\,,
\\[3pt]
\label{10072009-13}&& \ \ c_\nu \equiv \frac{\Gamma(\nu+\frac{d}{2})}{
\pi^{d/2} \Gamma(\nu)} \,.
\eeq
To be flexible, we use normalization factor $\sigma$ in \rf{19072009-09}. For
the case of scalar field, commonly used normalization in \rf{19072009-09} is
achieved by setting $\sigma=1$. Asymptotic behaviors of Green function
\rf{10072009-12} and solution \rf{19072009-09} are well known,
\beq
\label{10072009-14}
&& G_\nu(x,z) \ \ \ \stackrel{z \rightarrow 0}{\longrightarrow} \ \ \ z^{-\nu
+ \half} \delta^d(x)\,,
\\[5pt]
\label{man02-21072009-14}
&& \phi(x,z) \,\,\, \stackrel{z\rightarrow 0 }{\longrightarrow}\,\,\, z^{-\nu
+ \half} \sigma \phi_\sh(x)\,.
\eeq
From \rf{man02-21072009-14}, we see that our solution has indeed asymptotic
behavior corresponding to the shadow scalar field.

Using equations of motion \rf{19072009-06} in bulk action \rf{19072009-01}
with Lagrangian \rf{19072009-04} we obtain the
effective action given by%
\footnote{ Following commonly used setup, we consider solution of the
Dirichlet problem which tends to zero as $z\rightarrow \infty$. Therefore, in
\rf{19072009-07}, we ignore contribution to $S_\eff$ when $z=\infty$.}
\beq
\label{19072009-07} - S_\eff & = & \int d^dx\,  \LL_\eff\Bigr|_{z\rightarrow
0} \,,
\\[5pt]
\label{19072009-08} \LL_\eff & = & \half \phi \TT_{\nu -\half } \phi
\,.\qquad
\eeq

Plugging solution of the Dirichlet problem \rf{19072009-09} into
\rf{19072009-07}, \rf{19072009-08}, we obtain the effective action
\be \label{man02-21072009-15}
-S_\eff  = \nu c_\nu \sigma^2 \int d^dx_1d^dx_2
\frac{\phi_\sh(x_1)\phi_\sh(x_2)}{|x_{12}|^{2\nu + d}}\,.
\ee
Using the commonly used value of $\sigma$, $\sigma=1$, in
\rf{man02-21072009-15}, we obtain the properly normalized effective action
found in Refs.\cite{Gubser:1998bc,Freedman:1998tz}. Interesting novelty of
our computation of $S_\eff$ is that we use Fourier transform of the Green
function. Details of our computation may be found in Appendix C in
Ref.\cite{Metsaev:2009ym}.

\section{ AdS/CFT correspondence for spin-1 fields}\label{secAdS/CFTspin1}

We now discuss AdS/CFT correspondence for bulk spin-1 massive AdS field and
boundary spin-1 anomalous conformal current and shadow field. To this end we
are going to use CFT adapted gauge invariant Lagrangian and the modified
Lorentz gauge condition \cite{Metsaev:2009hp}%
\footnote{ For spin-1 massless field, the modified Lorentz gauge was found in
Ref.\cite{Metsaev:1999ui}, while for massless arbitrary spin field the
modified de Donder gauge was discovered in Ref.\cite{Metsaev:2008ks}.}.
Because our approach is closely related with gauge invariant approach to
massive field we start with brief review of the latter approach.

{\bf Gauge invariant approach to spin-1 massive field in $AdS_{d+1}$ space}.
In gauge invariant approach, spin-1 massive field is described by fields
\be \label{man11112010-13} \Phi^A\,, \qquad \Phi\,, \ee
which transform in the respective vector and scalar representations of
$so(d,1)$ algebra. In the Lorentzian signature, Lagrangian given by
\beq
e^{-1} \LL & = & - \frac{1}{4} F^{AB} F^{AB} -
\frac{1}{2} F^A F^A\,,
\nonumber\\[5pt]
&& F^{AB} \equiv \DD^A \Phi^B - \DD^B\Phi^A\,,
\nonumber\\[5pt]
\label{man11112010-10}  && F^A \equiv \DD^A \Phi+ m \Phi^A\,,
\eeq
is invariant under the gauge transformations
\be \label{man11112010-10a} \delta \Phi^A = \DD^A \Xi\,,\qquad \delta \Phi =
- m \Xi\,.\ee
Details of our notation may be found in Appendix \ref{app-01}. Lagrangian
\rf{man11112010-10} can be cast into the form which is more convenient for
our purposes,
\beq \label{man11112010-11}
&& e^{-1}\LL  =   \half \Phi^A (\DD^2 - m^2 +d)\Phi^A
\nonumber\\[5pt]
&& \hspace{1cm} + \,  \half \Phi(\DD^2 -m^2)\Phi + \half C_\st^2\,,
\\[5pt]
\label{man11112010-12} &&\hspace{1cm} C_\st \equiv \DD^C\Phi^C + m \Phi\,.
\eeq

\subsection{ CFT adapted gauge invariant approach to
spin-1 massive field in $AdS_{d+1}$}\label{sec09-a}

In our approach, the spin-1 massive AdS field is described by fields
\be \label{man11112010-14} \phi^a\,,\qquad \phi_{-1}\,,\qquad \phi_1\,,    \ee
which are the respective vector and scalar fields of the $so(d)$ algebra.
Fields in \rf{man11112010-14} are related by invertible transformation with
fields in \rf{man11112010-13} (see Appendix \ref{app-01}). CFT adapted
gauge invariant action and Lagrangian for field \rf{man11112010-14} take the
form,
\beq \label{09072009-02}
S & = & \int d^dx dz\,  \LL \,,
\\[5pt]
\label{man27102010-01}
\LL  & = &    \half |d\phi^a|^2 + \half |\TT_{\kappa-\half}\phi^a|^2
\nonumber\\[5pt]
& + & \half \sum_{\lambda =\pm 1}\Bigl(|d\phi_\lambda|^2 +
|\TT_{\kappa-\half+\lambda}\phi_\lambda|^2\Bigr) - \half C^2 \,, \qquad\quad
\\[5pt]
\label{11082009-03} && \hspace{-0.3cm} C \equiv \partial^a \phi^a + r_\zeta^{00}\TT_{\kappa +
\half}\phi_1 + r_z^{00}\TT_{-\kappa + \half}\phi_{-1}\,,\qquad
\eeq
where $\TT_\nu$ is given in \rf{09072009-04}, while $\kappa$ and $r_z^{00}$,
$r_\zeta^{00}$ are defined in \rf{man24102010-02n} and
\rf{man24102010-04} respectively. Lagrangian
\rf{man27102010-01} is invariant under gauge transformations
\beq
\label{11082009-04} && \delta \phi^a =\partial^a \xi\,,
\\[5pt]
\label{11082009-05} && \delta\phi_{-1} = r_z^{00} \TT_{\kappa-\half}\xi\,,
\\[5pt]
\label{11082009-06} && \delta\phi_1 = r_\zeta^{00} \TT_{-\kappa-\half}\xi\,,
\eeq
where $\xi$ is a gauge transformation parameter. Details of the derivation of
Lagrangian \rf{man27102010-01} from the one in \rf{man11112010-11} may be
found in Appendix \ref{app-01}.

Gauge invariant equations of motion obtained from Lagrangian \rf{man27102010-01}
take the form
\beq
&& \Box_{\kappa} \phi^a -\partial^a C  =0\,,
\nonumber\\[5pt]
\label{man27102010-02} && \Box_{\kappa-1} \phi_{-1} -
r_z^{00}\TT_{\kappa-\half}C =0\,,
\\[5pt]
&& \Box_{\kappa+1} \phi_1 - r_\zeta^{00}\TT_{-\kappa-\half}C =0\,,
\nonumber
\eeq
where the operator $\Box_\nu$ is given in \rf{09072009-09}.

{\bf Global AdS symmetries in CFT adapted approach}. General form of
realization of global symmetries for arbitrary spin AdS field was given in
\rf{conalggenlis01ads}-\rf{conalggenlis04ads}. All that is required to
complete description of the global symmetries is to find realization of the
operator $R_\smzero^a$ on space of gauge fields. For the case of spin-1
massive field, realization of the operator $R_\smzero^a$  on space of gauge
fields \rf{man11112010-14} is given by
\beq
&& R_\smzero^a\phi^b =  z \eta^{ab}
r_\zeta^{00} \phi_1 + z \eta^{ab} r_z^{00} \phi_{-1}\,,
\nonumber\\[5pt]
&& R_\smzero^a \phi_{-1} =- z r_z^{00} \phi^a \,,
\\[5pt]
&& R_\smzero^a \phi_1 = - z r_\zeta^{00} \phi^a \,.
\nonumber
\eeq

{\bf Modified Lorentz gauge}. Modified Lorentz gauge is defined to be
\be \label{09072009-11} C = 0\,,\qquad \hbox{modified Lorentz gauge} \,,\ee
where $C$ is given in \rf{11082009-03}. Using this gauge condition in
equations of motion \rf{man27102010-02} gives simple gauge fixed equations of
motion,
\beq
&& \Box_{\kappa} \phi^a =0\,,
\nonumber\\[-6pt]
\label{man27102010-07} &&
\\[-6pt]
&& \Box_{\kappa +\lambda} \phi_\lambda =0\,,\qquad \lambda =\pm 1\,.
\nonumber
\eeq
Thus, we see that the gauge fixed equations of motion are decoupled.

We note that the modified Lorentz gauge and gauge-fixed equations have leftover
on-shell gauge symmetry. Namely, modified Lorentz gauge \rf{09072009-11} and
gauge-fixed equations \rf{man27102010-07} are invariant under gauge
transformations given in \rf{11082009-04}-\rf{11082009-06} provided the gauge
transformation parameter satisfies the equation
\be \label{man05112010-05} \Box_\kappa\xi= 0\,.\ee

\subsection{ AdS/CFT correspondence for anomalous current
and normalizable modes of massive AdS field }\label{sec09-b}

We now ready to discuss AdS/CFT correspondence for spin-1 massive AdS field
and spin-1 anomalous conformal current. We begin with analysis of
normalizable solution of equations \rf{man27102010-07}. The normalizable
solution of equations \rf{man27102010-07} takes the form
\beq
&& \phi^a(x,z) = U_\kappa \phi_{\cur,0}^a(x)\,,
\nonumber\\[5pt]
\label{man01112010-28} && \phi_{-1}(x,z) = - U_{\kappa-1}\phi_{\cur,-1}(x)\,,
\\[5pt]
&& \phi_1(x,z) = U_{\kappa+1} \phi_{\cur,1}(x)\,,
\nonumber\\[5pt]
\label{man01112010-31} && \hspace{1cm} U_\nu \equiv h_\kappa \sqrt{zq}
J_\nu(zq) q^{-(\nu + \half)}\,,
\\[5pt]
&& \hspace{1cm} h_\kappa\equiv 2^\kappa\Gamma(\kappa+1)\,,\qquad  q^2\equiv
\Box\,. \qquad
\eeq
Note that we do not show explicitly dependence of $U_\nu$ on parameter
$\kappa$ \rf{man24102010-02n}. The asymptotic behavior of solution
\rf{man01112010-28} is given by
\beq
&& \phi^a(x,z) \ \ \stackrel{z\rightarrow 0}{\longrightarrow} \ \ z^{\kappa +
\half} \phi_{\cur,0}^a(x)\,,
\nonumber\\[5pt]
\label{man01112010-32} && \phi_{-1}(x,z) \ \ \stackrel{z\rightarrow
0}{\longrightarrow} \ \  - 2\kappa z^{\kappa - \half}  \phi_{\cur,-1}(x)\,,
\\[5pt]
&& \phi_1(x,z) \ \ \stackrel{z\rightarrow 0}{\longrightarrow} \ \
\frac{z^{\kappa + \frac{3}{2}}}{2(\kappa+1)} \phi_{\cur,1}(x)\,.
\nonumber
\eeq
From \rf{man01112010-32}, we see that $\phi_{\cur,0}^a$, $\phi_{\cur,\pm1}$
are indeed boundary values of the normalizable solution. In the r.h.s.
\rf{man01112010-28} we use the notation $\phi_{\cur,0}^a$, $\phi_{\cur,\pm1}$
since we are going to demonstrate that these boundary values are indeed the
gauge fields entering the gauge invariant formulation of the spin-1 anomalous
conformal current  in Sec.\ref{sec03}. Namely, one can prove the following
statements:

\noindent {\bf i}) For normalizable solution \rf{man01112010-28}, modified
Lorentz gauge condition \rf{09072009-11}  leads to differential constraint
\rf{man24102010-03} of the spin-1 anomalous conformal current.

\noindent {\bf ii}) {\it Leftover on-shell} gauge transformations
\rf{11082009-04}-\rf{11082009-06} of normalizable solution
\rf{man01112010-28} lead to gauge transformations
\rf{man24102010-06}-\rf{man24102010-08} of the spin-1 anomalous conformal
current%
\footnote{ Transformations given in \rf{11082009-04}-\rf{11082009-06} are
off-shell gauge transformations. Leftover on-shell gauge transformations are
obtained from \rf{11082009-04}-\rf{11082009-06} by using gauge transformation
parameter which satisfies equation \rf{man05112010-05}.}.

\noindent {\bf iii}) On-shell global $so(d,2)$ symmetries of the normalizable
modes of spin-1 massive $AdS_{d+1}$ field become global $so(d,2)$ conformal
symmetries of the spin-1 anomalous conformal current.

These statements can easily be proved by using the following relations for
the operator $U_\nu$:
\beq
\label{man01112010-35} && \TT_{\nu-\half} U_\nu  = U_{\nu-1}\,,
\\[5pt]
\label{man01112010-36} && \TT_{-\nu-\half} U_\nu  = - U_{\nu+1}\Box\,,
\\[5pt]
\label{man01112010-37} && \TT_{-\nu+\half} (zU_\nu)  = - z U_{\nu+1}\Box + 2
U_\nu \,,
\\[5pt]
\label{man01112010-37a1} && \Box_\nu (z U_{\nu+1}) = 2 U_\nu\,,
\eeq
which, in turn, can be obtained by using the following well-known identities
for the Bessel function:
\be \label{man01112010-38}
\TT_\nu J_{\nu } = J_{\nu-1}\,,
\qquad
\TT_{-\nu} J_{\nu } = - J_{\nu + 1}\,.
\ee

{\bf Matching of bulk modified Lorentz gauge and boundary constraint}. As an
illustration, we demonstrate how differential constraint for the anomalous conformal
current \rf{man24102010-03} can be obtained from modified Lorentz gauge
condition \rf{09072009-11}. To this end, adapting relations
\rf{man01112010-35} and \rf{man01112010-36} for the respective $\nu=\kappa+1$
and $\nu = \kappa-1$ we obtain the relations
\be
\label{man11112010-01} \TT_{\kappa+\half}U_{\kappa+1} = U_\kappa\,, \qquad
\TT_{-\kappa+\half}U_{\kappa-1} = - U_\kappa\Box\,.
\ee
Plugging solutions $\phi^a$, $\phi_{\pm1}$ \rf{man01112010-28} in $C$
\rf{11082009-03} and using \rf{man11112010-01} we obtain the relation
\be
\label{man01112010-38n} C = U_\kappa C_\cur\,,
\ee
where $C_\cur$ stands for left hand side of \rf{man24102010-03}. From
\rf{man01112010-38n}, we see that our modified Lorentz gauge condition $C=0$
\rf{09072009-11} leads indeed to differential constraint for the anomalous
conformal current \rf{man24102010-03}.

{\bf Matching of bulk and boundary gauge symmetries}. As the second
illustration, we demonstrate how gauge transformations of the anomalous
conformal current \rf{man24102010-06}-\rf{man24102010-08} can be obtained
from leftover on-shell gauge transformations of massive AdS field
\rf{11082009-04}-\rf{11082009-06}. To this end we note that the corresponding
normalizable solution of equation for gauge transformation parameter
\rf{man05112010-05} takes the form
\be
\label{man01112010-39}  \xi(x,z) = U_\kappa \xi_{\cur,0}(x)\,. \ee
Plugging $\phi^a$ \rf{man01112010-28} and $\xi$ \rf{man01112010-39} in
\rf{11082009-04}, we see that \rf{11082009-04} leads indeed to
\rf{man24102010-06}. To match boundary gauge transformation
\rf{man24102010-07} and bulk gauge transformation \rf{11082009-05} we plug
solution for $\xi$ \rf{man01112010-39} in bulk gauge transformation
\rf{11082009-05} and adapt relation \rf{man01112010-35} for $\nu=\kappa$ to
obtain
\beq \label{man11112010-02}
\delta\phi_{-1}(x,z) & = & r_z^{00} \TT_{\kappa-\half}
U_\kappa\xi_{\cur,0}(x)
\nonumber\\[5pt]
& = & U_{\kappa-1} r_z^{00} \xi_{\cur,0}(x)
\eeq
on the one hand. On the other hand, solution for $\phi_{-1}$
\rf{man01112010-28} implies
\be \label{man11112010-03}\delta \phi_{-1}(x,z) = - U_{\kappa-1} \delta
\phi_{\cur,-1}(x)\,.\ee
Comparing \rf{man11112010-02} and \rf{man11112010-03} we see that boundary
gauge transformation \rf{man24102010-07} and bulk gauge transformation
\rf{11082009-05} match. In the same way one can make sure that the remaining
boundary gauge transformation \rf{man24102010-08} and bulk gauge
transformation \rf{11082009-06} also match.

{\bf Matching of bulk and boundary global symmetries}. We note that
representation for generators given in
\rf{conalggenlis01ads}-\rf{conalggenlis04ads} is valid for gauge invariant
theory of AdS fields. This to say that our modified Lorentz gauge respects
the Poicar\'e and dilatation symmetries, but break the conformal boost
symmetries ($K^a$ symmetries). In other words, expressions for generators
$P^a$, $J^{ab}$ and $D$ given in
\rf{conalggenlis01ads}-\rf{conalggenlis03ads} are still valid for the
gauge-fixed AdS fields, while expression for the generator $K^a$
\rf{conalggenlis04ads} should be modified to restore $K^a$ symmetries for the
gauge-fixed AdS fields. Therefore, let us first to demonstrate matching of
the Poincar\'e and dilatation symmetries. What is required is to demonstrate
matching of the $so(d,2)$ algebra generators for bulk AdS field given in
\rf{conalggenlis01ads}-\rf{conalggenlis03ads} and the ones for boundary
conformal current given in \rf{conalggenlis01}-\rf{conalggenlis03}. As for
generators of the Poincar\'e algebra, $P^a$, $J^{ab}$, they already coincide
on both sides (see formulas \rf{conalggenlis01}, \rf{conalggenlis02} and the
respective formulas \rf{conalggenlis01ads},\rf{conalggenlis02ads}). Next,
consider the dilatation generator $D$. Here we need explicit form of solution
to bulk theory equations of motion given in \rf{man01112010-28}. Using the
notation $D_{_{AdS}}$ and $D_{_{CFT}}$ to indicate the respective
realizations of the dilatation generator $D$ on bulk field
\rf{conalggenlis03ads} and boundary current \rf{conalggenlis03}, we obtain
the relations
\beq
&& D_{_{AdS}} \phi^a(x,z) =  U_\kappa D_{_{CFT}} \phi_{\cur,0}^a(x)\,,
\nonumber\\[5pt]
&& D_{_{AdS}} \phi_{-1}(x,z) =  - U_{\kappa-1}
D_{_{CFT}}\phi_{\cur,-1}(x)\,,\qquad
\\[5pt]
&& D_{_{AdS}} \phi_1(x,z) =  U_{\kappa+1} D_{_{CFT}} \phi_{\cur,1}(x)\,,
\nonumber
\eeq
where $D_{_{CFT}}$ corresponding to $\phi_{\cur,0}^a$, $\phi_{\cur,-1}$,
$\phi_{\cur,1}$ can be obtained from \rf{conalggenlis03} and the respective
conformal dimensions \rf{man24102010-02}. Thus, the generators $D_{_{AdS}}$
and $D_{_{CFT}}$ also match.

We now turn to matching of the $K^a$ symmetries. As we have already said our
modified Lorentz gauge breaks the $K^a$ symmetries. To demonstrate this we
note that $K^a$ transformations of gauge fields \rf{man11112010-14} are given
by
\beq \label{man13112010-01}
&& K^a \phi^b = K_\Delta^a \phi^b + M^{abe} \phi^e
\nonumber\\[5pt]
&& \hspace{1cm} + \, z \eta^{ab} r_\zeta^{00} \phi_1 + z \eta^{ab} r_z^{00}
\phi_{-1} -  \half z^2 \partial^a \phi^b\,,\qquad\quad
\nonumber\\[5pt]
\label{man13112010-02} && K^a \phi_1 = K_\Delta^a \phi_1 - z r_\zeta^{00}
\phi^a  - \half z^2
\partial^a \phi_1\,,
\\[5pt]
\label{man13112010-03} && K^a \phi_{-1} = K_\Delta^a \phi_{-1} - z r_z^{00}
\phi^a  - \half z^2
\partial^a \phi_{-1}\,,
\nonumber
\eeq
where $K_\Delta^a$ and $M^{abc}$ are defined in
\rf{man19112010-01},\rf{man19112010-02}, while $\Delta$ is given in
\rf{conalggenlis03ads}. Using these transformation rules we find that $C$
\rf{11082009-03} transforms as
\be \label{man13112010-04}  K^a C = K_{\Delta+1}^aC - \half z^2 \partial^a C
- 2\phi^a\,,
\ee
i.e., we see that the modified Lorentz gauge condition $C=0$ is not invariant
under the $K^a$ transformations,
\be \label{man13112010-05} K^a C\left|_{C=0}^{\phantom{A}}\right. = -
2\phi^a\,.
\ee
This implies that generator $K^a$ given in \rf{conalggenlis04ads} should be
modified to restore the $K^a$ symmetries of the gauge-fixed AdS field theory.
To restore these broken $K^a$ symmetries we should, following standard
procedure, add compensating gauge transformations to maintain the $K^a$
symmetries. Thus, in order to find improved $K_\impr^a$ transformations of
the gauge-fixed AdS fields \rf{man11112010-14} we start with the generic
global $K^a$ transformations \rf{man13112010-02} supplemented by the
appropriate compensating gauge transformations
\beq
&& K_\impr^a\phi^b = K^a\phi^b +
\partial^b\xi^{K^a}\,,
\nonumber\\[5pt]
\label{man13112010-06} && K_\impr^a\phi_{-1} = K^a\phi_{-1} + r_z^{00}
\TT_{\kappa-\half}\xi^{K^a}\,,
\\[5pt]
&& K_\impr^a\phi_1 = K^a\phi_1 + r_\zeta^{00} \TT_{-\kappa-\half}\xi^{K^a}\,,
\nonumber
\eeq
where $\xi^{K^a}$ stands for parameter of the compensating gauge
transformations. Computing $K_\impr^a$ transformation of $C$
\be \label{man13112010-09} K_\impr^a C = K_{\Delta+1}^a C - \half z^2
\partial^a C - 2\phi^a + \Box_\kappa \xi^{K^a}\,,
\ee
and requiring the $K_\impr^a$ transformation to maintain the gauge condition
$C=0$,
\be \label{man13112010-10} K_\impr^a C\left|_{C=0}^{\phantom{A}}\right. =
0\,, \ee
we get the equation for $\xi^{K^a}$
\be \label{man13112010-11} \Box_\kappa \xi^{K^a} - 2\phi^a =0 \,.\ee
Thus, we obtain the non-homogeneous second-order differential equation for
the compensating gauge transformation parameter $\xi^{K^a}$. Plugging
normalizable solution \rf{man01112010-28} in \rf{man13112010-11} we obtain
the equation
\be \label{man13112010-12} \Box_\kappa \xi^{K^a}(x,z) = 2U_\kappa
\phi_{\cur,0}^a(x)\,.\ee
Using \rf{man01112010-37a1}, solution to equation \rf{man13112010-12} is
easily found to be
\be
\label{man13112010-13}  \xi^{K^a}(x,z) = z U_{\kappa+1}\phi_{\cur,0}^a(x)\,.
\ee
Plugging \rf{man01112010-28} and \rf{man13112010-13} in \rf{man13112010-06},
we make sure that improved $K_\impr^a$ transformations lead to the conformal
boost transformations for the spin-1 anomalous conformal current given in
\rf{04092008-01},\rf{conalggenlis04} with operator $R^a$ defined in
\rf{man24102010-08a}.

\subsection{ AdS/CFT correspondence for anomalous shadow field
and non-normalizable mode of massive AdS field.}\label{sec09-c}

We proceed to discussion of AdS/CFT correspondence for bulk spin-1 massive
AdS field and boundary spin-1 anomalous shadow field.

{\bf Matching of effective action and boundary two-point vertex}. In order to
find bulk effective action $S_\eff$ we should, following the standard
strategy, solve bulk equations of motion with the Dirichlet problem
corresponding to the boundary anomalous shadow field and plug the solution
into bulk action. Using gauge invariant equations of motion
\rf{man27102010-02} in bulk action \rf{09072009-02}, we obtain the following
effective action:
\beq
\label{man27102010-05} S_\eff & = & - \int d^dx\,  \LL_\eff\Bigr|_{z\rightarrow
0} \,,
\\[5pt]
\label{man27102010-06} \LL_\eff & = & \half \phi^a \TT_{\kappa -\half } \phi^a +
\half \sum_{\lambda=\pm 1} \phi_\lambda\TT_{\kappa -\half+\lambda } \phi_\lambda
\nonumber\\[5pt]
& - & \half (r_z^{00}\phi_{-1} + r_\zeta^{00}\phi_1) C  \,.
\eeq

As we have already seen, use of the modified Lorentz gauge considerably
simplifies the equations of motion. Now, using modified Lorentz gauge
\rf{09072009-11} in \rf{man27102010-06}, we obtain
\be \label{man27102010-10}
\LL_\eff\Bigr|_{C=0} =  \half \phi^a \TT_{\kappa -\half } \phi^a + \half
\sum_{\lambda =\pm 1} \phi_\lambda \TT_{\kappa -\half+\lambda } \phi_\lambda
\,,
\ee
i.e. we see that $\LL_\eff$ is also simplified. In order to find $S_\eff$ we
should solve gauge fixed equations of motion \rf{man27102010-07} with the
Dirichlet problem corresponding to the boundary anomalous shadow field and
plug the solution into \rf{man27102010-10}. We now discuss solution to
equations of motion \rf{man27102010-07}.

Because gauge fixed equations of motion \rf{man27102010-07} are similar to
the ones for scalar AdS field \rf{19072009-06} we can simply apply result
in Sec. \ref{secAdS/CFT}. This is to say that solution of equations
\rf{man27102010-07} with the Dirichlet problem corresponding to the spin-1
anomalous shadow field takes the form
\beq
\phi^a(x,z)\ & = & \sigma_{1,0} \int d^dy\, G_{\kappa}(x-y,z)
\phi_{\sh,0}^a(y)\,,\
\nonumber\\[5pt]
\label{man27102010-11} \phi_{-1}(x,z)\! & = &  \sigma_{0,-1}\!\!\! \int
d^dy\, G_{\kappa-1}(x-y,z) \phi_{\sh,1}(y)\,,\qquad
\\[5pt]
\phi_1(x,z)\ & = &  \sigma_{0,1}\!\!\! \int d^dy\, G_{\kappa+1}(x-y,z)
\phi_{\sh,-1}(y)\,,\qquad\
\nonumber\\[5pt]
\label{man27102010-13}&& \hspace{-1cm}\sigma_{1,0} \equiv 1\,,
\\[5pt]
\label{man27102010-14} && \hspace{-1cm}\sigma_{0,-1} \equiv -\frac{1}{2(\kappa-1)}\,,
\qquad \sigma_{0,1} \equiv 2\kappa\,,
\eeq
where the Green function is given in \rf{10072009-12}.

Using asymptotic behavior of the Green function $G_\nu$ \rf{10072009-14}, we
find the asymptotic behavior of our solution
\beq
&& \phi^a(x,z) \,\,\, \stackrel{z\rightarrow 0 }{\longrightarrow}\,\,\,
z^{-\kappa + \half} \phi_{\sh,0}^a(x)\,,
\nonumber\\[5pt]
\label{man01112010-04} && \phi_{-1}(x,z) \,\,\, \stackrel{z\rightarrow 0
}{\longrightarrow}\,\,\,  - \frac{z^{-\kappa + \frac{3}{2} }}{2(\kappa-1)}
\phi_{\sh,1}(x)\,,
\\[5pt]
&& \phi_1(x,z) \,\,\, \stackrel{z\rightarrow 0 }{\longrightarrow}\,\,\,
2\kappa z^{-\kappa -\half} \phi_{\sh,-1}(x)\,.
\nonumber
\eeq
From these expressions, we see that our solution has indeed asymptotic
behavior corresponding to the spin-1 anomalous shadow field. Note that
because the solution has non-integrable asymptotic behavior
\rf{man01112010-04}, such solution is referred to as the non-normalizable
solution in the literature.

We now explain the choice of the normalization factors $\sigma_{1,0}$,
$\sigma_{0,\pm 1}$ in \rf{man27102010-13}, \rf{man27102010-14}. The choice of
$\sigma_{1,0}$ is a matter of convention. Following commonly used convention,
we set this normalization factor to be equal to 1. The remaining
normalization factors $\sigma_{0,\pm 1}$ are then determined uniquely by
requiring that the modified Lorentz gauge condition for the spin-1 massive
AdS field \rf{09072009-11} be amount to the differential constraint for the
spin-1 anomalous shadow field \rf{man24102010-22}. With the choice made in
\rf{man27102010-13},\rf{man27102010-14} we find the relations
\beq
&& \hspace{-1.1cm} \partial^a\phi^a  =  \int d^dy\,
G_\kappa(x-y,z)\partial^a\phi_{\sh,0}^a(y)\,,
\nonumber\\[5pt]
\label{man27102010-16} && \hspace{-1.1cm} \TT_{-\kappa+\half} \phi_{-1} =
\!\!\int\! d^dy\, G_\kappa(x-y,z) \phi_{\sh,1}(y),
\\[5pt]
&& \hspace{-1.1cm} \TT_{\kappa+\half} \phi_1 = \int d^dy\, G_\kappa(x-y,z)
\Box \phi_{\sh,-1}(y)\,.
\nonumber
\eeq
From these relations and \rf{11082009-03}, we see that our choice of
$\sigma_{1,\pm 1}$ \rf{man27102010-14} allows us to match modified Lorentz
gauge for the spin-1 massive AdS field \rf{09072009-11} and differential
constraint for the spin-1 anomalous shadow field given in
\rf{man24102010-22}. We note the helpful relations for the Green function
which we use for the derivation of relations \rf{man27102010-16},
\beq
&& \TT_{-\kappa+\half}G_{\kappa-1} = - 2(\kappa-1) G_\kappa,,
\nonumber\\[-6pt]
\label{man01112010-24} &&
\\[-6pt]
&& \TT_{\kappa+\half}G_{\kappa+1} = \frac{1}{2\kappa} \Box G_\kappa\,,
\nonumber
\eeq
where $G_\nu \equiv G_\nu(x-y,z)$.

All that remains to obtain $S_\eff$ is to plug solution of the Dirichlet
problem for AdS field \rf{man27102010-11} into \rf{man27102010-05},
\rf{man27102010-10}. Using general formula given in \rf{man02-21072009-15},
we obtain
\be \label{man08112010-01}
-S_\eff  =  2\kappa c_\kappa \Gamma \,,
\ee
where $\kappa$ and $c_\kappa$ are defined in
\rf{man24102010-02n},\rf{10072009-13} respectively and $\Gamma$ is gauge
invariant two-point vertex of the spin-1 anomalous shadow field given in
\rf{man05112010-18},\rf{man05112010-19}.

Thus we see that {\it imposing the modified Lorentz gauge on the spin-1
massive AdS field and computing the bulk action on the solution of equations
of motion with the Dirichlet problem corresponding to the boundary anomalous
shadow field we obtain the gauge invariant two-point vertex of the spin-1
anomalous shadow field}.

Because in the literature $S_\eff$ is expressed in terms of two-point vertex
taken in the Stueckelberg gauge frame, $\Gamma^{{\rm stand}}$
\rf{man05112010-22}, we use \rf{man05112010-21} and represent our result
\rf{man08112010-01} as
\be \label{man08112010-02}
-S_\eff  =  \frac{\kappa(2\kappa+d)}{2\kappa+d-2} c_\kappa \Gamma^{{\rm stand}} \,.
\ee
This relation was obtained in Ref.\cite{Mueck:1998iz}. The fact that $S_\eff$
is proportional to $\Gamma^{{\rm stand}}$ is expected because of the
conformal symmetry, but for the systematical study of AdS/CFT
correspondence it is important to know the normalization factor in front of
$\Gamma^{{\rm stand}}$ \rf{man08112010-02}. Our normalization factor
coincides with the one found in Ref.\cite{Mueck:1998iz}%
\footnote{ Computation of $S_\eff$ for spin-1 massless field may be found in
Ref.\cite{Freedman:1998tz} and, in the framework of our approach, in
Ref.\cite{Metsaev:2009ym}.}.

Note that we have obtained more general relation given in
\rf{man08112010-01}, while relation \rf{man08112010-02}  is obtained from
\rf{man08112010-01} by using the Stueckelberg gauge frame. Attractive feature
of our approach is that it provides possibility to use other gauge conditions
which might be preferable in certain applications. This is to say that, in
the light-cone gauge frame, relation \rf{man08112010-01} takes the form
\be \label{2m26122010-02}
-S_\eff  =  2\kappa c_\kappa \Gamma^{({\rm l.c.})} \,.
\ee
Note that transformation of relation \rf{man08112010-02} to the one in
\rf{2m26122010-02} requires cumbersome computations because the Stueckelberg
gauge frame removes the scalar field entering the light-cone gauge frame (see
Secs.\ref{sec004-b} and \ref{sec004-c}). {\it It is relation
\rf{2m26122010-02} that seems to be most suitable for the the study of
duality of the light-cone gauge Green-Schwarz AdS superstring and the
corresponding boundary gauge theory}.

{\bf Matching of bulk and boundary gauge symmetries}. Modified Lorentz gauge
\rf{09072009-11} and gauge-fixed equations \rf{man27102010-07} are invariant
under gauge transformations given in \rf{11082009-04}-\rf{11082009-06}
provided the gauge transformation parameter satisfies equation
\rf{man05112010-05}. The non-normalizable solution to this equation is given
by
\be \label{man01112010-26} \xi(x,z)  =  \int d^dy\, G_\kappa(x-y,z)\xi_\sh(y)\,.\ee
We now note that, on the one hand, plugging \rf{man01112010-26} in
\rf{11082009-04}-\rf{11082009-06} and using relations \rf{man01112010-24} we
represent on-shell gauge transformations of $\phi^a(x,z)$, $\phi_{-1}(x,z)$
and $\phi_1(x,z)$ as
\beq
&& \delta \phi^a  =  \int d^dy\, G_\kappa(x-y,z)\partial^a\xi_\sh(y)\,,
\nonumber\\[5pt]
\label{man27102010-19} && \delta \phi_{-1} = \frac{r_z^{00}}{2(\kappa-1)}
\int d^dy\, G_{\kappa-1}(x-y,z) \Box \xi_\sh(y)\,,\qquad\quad
\\[5pt]
&& \delta \phi_1 = -2\kappa r_\zeta^{00} \int d^dy\, G_{\kappa+1}(x-y,z)
\xi_\sh(y)\,.\qquad
\nonumber
\eeq
On the other hand, relations \rf{man27102010-11} imply
\beq
&& \hspace{-0.4cm}\delta\phi^a(x,z)  =  \sigma_{1,0} \int d^dy\,
G_{\kappa}(x-y,z) \delta\phi_{\sh,0}^a(y)\,,
\nonumber\\[5pt]
\label{man27102010-22}
&& \hspace{-0.4cm} \delta\phi_{-1}(x,z)  =  \sigma_{1,-1}\!\!\! \int d^dy\,
G_{\kappa-1}(x-y,z) \delta\phi_{\sh,1}(y)\,,\qquad\quad
\\[5pt]
&& \hspace{-0.4cm} \delta\phi_1(x,z)  = \sigma_{1,1}\!\!\! \int d^dy\,
G_{\kappa+1}(x-y,z) \delta\phi_{\sh,-1}(y)\,.\qquad
\nonumber
\eeq
Comparing \rf{man27102010-19} with \rf{man27102010-22} we see that the
on-shell leftover gauge symmetries of solution of the Dirichlet problem for
spin-1 massive AdS field amount to gauge symmetries of the spin-1 anomalous
shadow field \rf{man24102010-23}-\rf{man24102010-25}.

{\bf Matching of bulk and boundary global symmetries}. The matching can be
demonstrated by following the procedure we used for the spin-1 anomalous
current in Sec.\ref{sec09-b}. Therefore to avoid repetitions we briefly
discuss some necessary details. Matching of bulk and boundary Poinca\'e
symmetries is obvious. Using conformal dimensions for spin-1 anomalous shadow
field given in \rf{man24102010-21}, solution for bulk fields in
\rf{man27102010-11}, and bulk dilatation operator \rf{conalggenlis03ads} we
make sure that dilatation bulk and boundary symmetries also match. In order
to match $K^a$ symmetries we consider improved $K_\impr^a$ transformations
with compensating gauge transformation parameters satisfying equations
\rf{man13112010-11}. Using the relation for the Green function
\be \label{2m26122010-04}  \Box_\nu (zG_{\nu-1}) = - 4(\nu-1) G_\nu \,, \ee
it is easy to see that solution to equation \rf{man13112010-11} with $\phi^a$
as in \rf{man27102010-11} is given by
\beq \label{2m26122010-05}
&& \hspace{-1cm} \xi^{K^a}(x,z) =  z \sigma_{1,0}^\xi\! \int\!\! d^dy\,
G_{\kappa-1}(x-y,z) \phi_{\sh,0}^a(y)\,,\qquad
\\[3pt]
&& \hspace{1cm} \sigma_{1,0}^\xi \equiv -\frac{1}{2(\kappa-1)}\,.
\eeq
Using \rf{man27102010-11} and \rf{2m26122010-05} in \rf{man13112010-06}, we
make sure that improved bulk $K_\impr^a$ symmetries amount to $K^a$
symmetries of the spin-1 anomalous shadow field given in \rf{conalggenlis04}
and \rf{man24102010-26n}.

To summarize, we note that it is {\it matching of the bulk on-shell leftover
gauge symmetries of the solution to Dirichlet problem and bulk global
symmetries and the respective boundary gauge symmetries of the anomalous
shadow field and boundary global symmetries that explains why the effective
action coincides with the gauge invariant two-point vertex for the boundary
anomalous shadow field (see \rf{man08112010-01})}.

\section{ AdS/CFT correspondence for spin-2 fields.}\label{secAdS/CFTspin2}

Before discussing AdS/CFT correspondence for spin-2 massive AdS field and
spin-2 anomalous conformal current and shadow field we present our CFT
adapted gauge invariant approach to spin-2 massive AdS field. Because our
approach is closely related with gauge invariant approach to massive field we
start with brief review of the latter approach.

{\bf Gauge invariant approach to spin-2 massive field in $AdS_{d+1}$ space}.
In gauge invariant approach, spin-2 massive field is described by gauge
fields
\be \label{man11112010-15} \Phi^{AB}\,,\qquad \Phi^A\,, \qquad \Phi\,, \ee
which transform in the respective rank-2 tensor, vector and scalar
representations of $so(d,1)$ algebra. In Lorentzian signature, Lagrangian
found in Ref.\cite{Zinoviev:2001dt} takes the form%
\footnote{ Recent interesting discussion of massive AdS fields may be found
in \cite{Ponomarev:2010st}.}
\beq \label{27012009-01x}
\frac{1}{e}\LL & = & \frac{1}{4} \Phi^{AB}E_{_{EH}}\Phi^{AB} + \half \Phi^A
E_{_{Max}}\Phi^A + \half \Phi \DD^2 \Phi
\nonumber\\[3pt]
& + & m \Phi^A (\DD^B \Phi^{BA}  - \DD^A \Phi^{BB}) + f\Phi \DD^A\Phi^A
\nonumber\\[3pt]
& - & \frac{m^2-2}{4} \Phi^{AB} \Phi^{AB} +  \frac{m^2+d-2}{4} \Phi^{AA}
\Phi^{BB}
\nonumber\\[3pt]
& + & \frac{fm}{2} \Phi^{AA} \Phi - \frac{d}{2}\Phi^A\Phi^A +
\frac{(d+1)m^2}{2(d-1)}\Phi^2 \,,
\\[5pt]
\label{man11112010-20a} && \hspace{1cm} f \equiv \left(\frac{2d}{d-1}m^2 +
2d\right)^{1/2}\,,
\eeq
where the respective second-derivative Einstein-Hilbert and Maxwell operators
$E_{_{EH}}$, $E_{_{Max}}$ are given by
\beq
&& \hspace{-0.7cm} E_{_{EH}}\Phi^{AB} = \DD^2 \Phi^{AB} - \DD^A\DD^C\Phi^{CB}
- \DD^B \DD^C \Phi^{CA}
\nonumber\\[5pt]
&& + \, \DD^A \DD^B \Phi^{CC}  +  \eta^{AB}( \DD^C \DD^E\Phi^{CE} - \DD^2
\Phi^{CC}) \,,\qquad
\nonumber\\[5pt]
&& \hspace{-0.7cm} E_{_{Max}}\Phi^A = \DD^2 \Phi^A - \DD^A \DD^B \Phi^B \,.
\eeq
Lagrangian \rf{27012009-01x} is invariant under gauge transformations
\beq
&& \hspace{-0.7cm} \delta \Phi^{AB} = \DD^A \Xi^B + \DD^B \Xi^A + \frac{2m
}{d-1} \eta^{AB} \Xi\,,
\nonumber\\[5pt]
\label{man11112010-18} && \hspace{-0.7cm} \delta \Phi^A = \DD^A \Xi - m \Xi^A
\,,
\\[5pt]
&& \hspace{-0.7cm} \delta \Phi  = - f \Xi\,,
\nonumber
\eeq
where $\Xi^A$, $\Xi$ are gauge transformation parameters. In
Ref.\cite{Metsaev:2009hp}, we found new representation for Lagrangian
\rf{27012009-01x},
\beq \label{man11112010-21}
\frac{1}{e}\LL & = & \frac{1}{4} \Phi^{AB} (\DD^2 -m^2 +2) \Phi^{AB}
\nonumber\\[5pt]
& - & \frac{1}{8} \Phi^{AA} (\DD^2 -m^2 -2d+4) \Phi^{BB}
\nonumber\\[3pt]
& + & \half \Phi^A (\DD^2 -m^2-d)\Phi^A+ \half \Phi (\DD^2 -m^2-2d)\Phi
\nonumber\\[3pt]
& + & \half C_\st^A C_\st^A + \half C_\st^2\,,
\\[5pt]
&& C_\st^A = \DD^B \Phi^{BA} - \half \DD^A \Phi^{BB} + m \Phi^A\,,
\nonumber\\[-6pt]
\label{man11112010-22} &&
\\[-6pt]
&& C_\st  = \DD^A\Phi^A + \frac{m}{2}\Phi^{AA} + f \Phi\,.
\nonumber
\eeq
From \rf{man11112010-21}, we see that it is the use of quantities $C_\st^A$
and $C_\st$ that simplifies the structure of the gauge invariant Lagrangian.
We note also that the relations $C_\st^A=0$, $C_\st = 0$ define standard de
Donder gauge condition for the spin-2 massive field%
\footnote{ Recent discussion of the {\it standard} de Donder-Feynman gauge
for massless fields may be found in
Refs.\cite{Guttenberg:2008qe,Manvelyan:2008ks,Fotopoulos:2009iw}. To our
knowledge explicit form of $C_\st^A$, $C_\st$ \rf{man11112010-22} has not
been discussed in the earlier literature.}.

{\bf Interrelation of gauge invariant Lagrangian and Pauli-Fierz Lagrangian}.
As is well known, the spin-2 massive AdS field can be described by the
Pauli-Fierz Lagrangian given by
\beq \label{man11112010-24}
\frac{1}{e} \LL_{_{PF}}  & = &  \frac{1}{4} \Phi_{_{PF}}^{AB}
(E_{_{EH}}\Phi_{_{PF}})^{AB}
-  \frac{m^2-2}{4} \Phi_{_{PF}}^{AB} \Phi_{_{PF}}^{AB}
\nonumber\\[5pt]
& + & \frac{m^2+d-2}{4} \Phi_{_{PF}}^{AA} \Phi_{_{PF}}^{BB}\,,
\eeq
where $\Phi_{_{PF}}^{AB}$ is rank-2 tensor field of $so(d,1)$ algebra.
Pauli-Fierz Lagrangian can be obtained from gauge invariant Lagrangian
\rf{27012009-01x} in obvious way. Namely, gauge transformations
\rf{man11112010-18} allow us to gauge away the fields $\Phi^A$ and $\Phi$.
Doing so and  identifying rank-2 tensor field in \rf{man11112010-15} with
$\Phi_{_{PF}}^{AB}$, we get the Pauli-Fierz Lagrangian from gauge invariant
Lagrangian \rf{27012009-01x},
\be \label{man11112010-25} \LL_{_{PF}} = \LL|_{ \Phi^{AB} \equiv
\Phi_{_{PF}}^{AB},\, \Phi^A = 0,\,\Phi=0 }\,. \ee

For the case of flat space, it is well known that the gauge invariant
Lagrangian can be obtained from the Pauli-Fierz Lagrangian. It turns out that
this interrelation is still to be valid in AdS space too. Namely, introducing
the following representation of the Pauli-Fierz field in terms of gauge
fields \rf{man11112010-15}
\beq \label{man11112010-26}
\Phi_{_{PF}}^{AB} & = & \Phi^{AB} + \frac{1}{m} (\DD^A \Phi^B +  \DD^B
\Phi^A)
\nonumber\\[5pt]
& + & \frac{2}{mf} \DD^A \DD^B\Phi + \frac{2m}{(d-1)f} \eta^{AB} \Phi\,,
\qquad \eeq
and plugging such $\Phi_{_{PF}}^{AB}$ \rf{man11112010-26} into Pauli-Fierz
Lagrangian \rf{man11112010-21}, we obtain gauge invariant
Lagrangian \rf{27012009-01x}%
\footnote{To our knowledge formula \rf{man11112010-26} is new and has not
been discussed in the earlier literature. For $4d$ flat space, formula
\rf{man11112010-26} was given in Ref.\cite{Schwinger:1970xc}, while for flat
space with $d>4$ in Ref.\cite{Metsaev:2008fs}.}.

\subsection{CFT adapted gauge invariant approach to spin-2 massive field in $AdS_{d+1}$}

We now discuss our CFT adapted approach to spin-2 massive AdS field. For
details of the derivation of the CFT adapted gauge invariant Lagrangian, see
Appendix \ref{app-02}.

In our approach, the spin-2 massive field is described by the gauge fields
\beq \label{man12112010-05}
& \phi^{ab}  &
\nonumber\\[5pt]
& \phi_{-1}^a\qquad \phi_1^a  &
\\[5pt]
& \phi_{-2} \qquad \phi_0  \qquad \phi_2 &
\nonumber
\eeq
The fields $\phi^{ab}$, $\phi_{\pm 1}^a$ and $\phi_0$, $\phi_{\pm 2}$ are the
respective rank-2 tensor, vector and scalar fields of the $so(d)$ algebra.
The CFT adapted gauge invariant Lagrangian for these fields takes the form
\cite{Metsaev:2009hp}
\beq \label{man29102010-11}
\LL & = &  \frac{1}{4} |d\phi^{ab}|^2 -\frac{1}{8} |d\phi^{aa}|^2 +
\frac{1}{4}|\TT_{\kappa-\half}\phi^{ab}|^2-
\frac{1}{8}|\TT_{\kappa-\half}\phi^{aa}|^2
\nonumber\\[5pt]
& + &  \half \sum_{\lambda=\pm 1} \Bigl( |d\phi_\lambda^a|^2 +
|\TT_{\kappa-\half+\lambda }\phi_\lambda^a|^2\Bigr)
\nonumber\\[5pt]
& + &  \half \sum_{\lambda=0,\pm 2} \Bigr( |d\phi_\lambda|^2 +
|\TT_{\kappa - \half +\lambda}\phi_\lambda|^2\Bigl)
\nonumber\\[5pt]
& - & \half C^aC^a - \half C_1 C_1 - \half C_{-1}C_{-1}\,,
\eeq
where we use the notation
\beq
C^a  & \equiv & \partial^b \phi^{ab} -\half \partial^a\phi^{bb}
\nonumber\\[5pt]
& + &  r_z^{00} \TT_{-\kappa +\half}
\phi_{-1}^a + r_\zeta^{00} \TT_{\kappa+ \half} \phi_1^a  \,,
\nonumber\\[5pt]
\label{man05112010-01} C_1\, & \equiv &  \partial^a \phi_1^a -\half
r_\zeta^{00} \TT_{- \kappa - \half } \phi^{aa}
\\[5pt]
& +  & r_z^{10} \TT_{- \kappa -\half } \phi_0
+ \sqrt{2} r_\zeta^{10} \TT_{\kappa +\frac{3}{2}} \phi_2\,,
\nonumber\\[5pt]
C_{-1}\! & \equiv &  \partial^a \phi_{-1}^a -\half r_z^{00} \TT_{ \kappa -
\half } \phi^{aa}
\nonumber\\[5pt]
& + & \sqrt{2} r_z^{01} \TT_{-\kappa +\frac{3}{2}} \phi_{-2}
+ r_\zeta^{01} \TT_{\kappa -\half} \phi_0 \,,
\nonumber
\eeq
and $\TT_\nu$ is given in \rf{09072009-04}, while $\kappa$ and $r_z^{mn}$,
$r_\zeta^{mn}$ are defined in \rf{man24102010-39} and
\rf{man25102010-01} respectively. Lagrangian
\rf{man29102010-11} is invariant under the gauge transformations
\beq
&& \delta\phi^{ab} = \partial^a\xi^b + \partial^b
\xi^a
\nonumber\\[5pt]
&& \hspace{0.7cm} +\,
\frac{2r_\zeta^{00}}{d-2}\eta^{ab} \TT_{\kappa + \half} \xi_{1} +
\frac{2r_z^{00}}{d-2}\eta^{ab} \TT_{-\kappa + \half} \xi_{-1}\,,
\nonumber\\[5pt]
&& \delta\phi_{-1}^a =\partial^a\xi_{-1} + r_z^{00}
\TT_{\kappa -\half} \xi^a\,,
\nonumber\\[5pt]
\label{man05112010-03a1} && \delta\phi_1^a =\partial^a \xi_1 + r_\zeta^{00}
\TT_{-\kappa -\half} \xi^a\,,
\\[5pt]
&& \delta\phi_{-2} = \sqrt{2} r_z^{01} \TT_{\kappa -
\frac{3}{2}} \xi_{-1}\,,
\nonumber\\[5pt]
&& \delta\phi_0 = r_z^{10} \TT_{\kappa + \half}
\xi_1 + r_\zeta^{01} \TT_{-\kappa +\half } \xi_{-1}
\,,\qquad
\nonumber\\[5pt]
&& \delta\phi_2 = \sqrt{2} r_\zeta^{10} \TT_{-
\kappa -\frac{3}{2}} \xi_1\,,
\nonumber
\eeq
where $\xi^a$, $\xi_{\pm 1}$ are gauge transformation parameters.

Gauge invariant equations of motion obtained from Lagrangian \rf{man29102010-11}
take the form
\beq
&& \hspace{-0.3cm} \Box_\kappa\phi^{ab} - \partial^a C^b - \partial^b C^a
\nonumber\\[5pt]
&& - \frac{2r_z^{00}\eta^{ab}}{d-2}\TT_{-\kappa+\half}C_{-1}
- \frac{2r_\zeta^{00}\eta^{ab}}{d-2} \TT_{\kappa+\half}C_1 =0 \,,\qquad\quad
\nonumber\\[5pt]
&& \hspace{-0.3cm} \Box_{\kappa-1}\phi_{-1}^a -
\partial^a C_{-1} - r_z^{00} \TT_{\kappa-\half}C^a = 0 \,,
\nonumber\\[5pt]
\label{man14112010-01} && \hspace{-0.3cm} \Box_{\kappa+1}\phi_1^a -
\partial^a C_1 - r_\zeta^{00} \TT_{-\kappa-\half}C^a = 0 \,,
\\[5pt]
&& \hspace{-0.3cm} \Box_{\kappa-2}\phi_{-2}
-\sqrt{2}r_z^{01}\TT_{\kappa-\frac{3}{2}}C_{-1} = 0 \,,
\nonumber\\[5pt]
&& \hspace{-0.3cm} \Box_\kappa\phi_0  - r_\zeta^{01}\TT_{-\kappa+\half}C_{-1}
- r_z^{10}\TT_{\kappa+\half} C_1 = 0 \,,
\nonumber\\[5pt]
&& \hspace{-0.3cm} \Box_{\kappa+2}\phi_2
-\sqrt{2}r_\zeta^{10}\TT_{-\kappa-\frac{3}{2}}C_1 = 0 \,,
\nonumber
\eeq
where $\Box_\nu$ is defined in \rf{09072009-09}. We see that the gauge
invariant equations of motion are coupled.

{\bf Global AdS symmetries}. We now discuss realization of the global AdS
symmetries on space of gauge fields \rf{man12112010-05}. The realization of
the global $AdS$ symmetries is already given in
\rf{conalggenlis01ads}-\rf{14092008-08}. All that remains to complete the
description of these symmetries is to find realization of the operator
$R_\smzero^a$ on space of gauge fields \rf{man12112010-05}. Action of the
operator $R_\smzero^a$ on space of gauge fields \rf{man12112010-05} is found
to be,
\beq
&& \hspace{-1cm} R_\smzero^a\phi^{bc} =  zr_\zeta^{00} ( \eta^{ab}\phi_1^c +
\eta^{ac}\phi_1^b - \frac{2\eta^{bc}}{d-2} \phi_1^a)
\nonumber\\[3pt]
&&  + \, zr_z^{00} ( \eta^{ab}\phi_{-1}^c + \eta^{ac}\phi_{-1}^b -
\frac{2\eta^{bc}}{d-2}  \phi_{-1}^a),
\nonumber\\[5pt]
&& \hspace{-1cm}  R_\smzero^a\phi_1^b  =  -z r_\zeta^{00}\phi^{ab} + z
\eta^{ab}( \sqrt{2}r_\zeta^{10}\phi_2 + r_z^{10}\phi_0 ),
\nonumber\\[5pt]
&& \hspace{-1cm} R_\smzero^a\phi_{-1}^b  =   -z r_z^{00}\phi^{ab} +
z\eta^{ab}( \sqrt{2}r_z^{01}\phi_{-2} +  r_\zeta^{01}\phi_0),
\nonumber\\
&&\\[-5pt]
&& \hspace{-1cm} R_\smzero^a\phi_2  =  -z\sqrt{2} r_\zeta^{10}\phi_1^a\,,
\nonumber\\[5pt]
&& \hspace{-1cm} R_\smzero^a\phi_0  =  -zr_z^{10}\phi_1^a -z
r_\zeta^{01}\phi_{-1}^a\,,
\nonumber\\[5pt]
&& \hspace{-1cm} R_\smzero^a \phi_{-2} =   -z\sqrt{2} r_z^{01}\phi_{-1}^a\,.
\nonumber
\eeq

{\bf Modified de Donder gauge}. Modified de Donder
gauge is defined to be
\be
\label{man01112010-16} C^a=0,\quad C_{-1} = 0,\quad C_1=0,\quad
\hbox{modified de Donder gauge},
\ee
where $C^a$, $C_{\pm1}$ are given in \rf{man05112010-01}. Using this gauge in
equations of motion \rf{man14112010-01} gives the surprisingly simple gauge
fixed equations of motion,
\beq
\label{man01112010-01} && \Box_\kappa\phi^{ab}=0\,,
\nonumber\\[5pt]
\label{man01112010-02}&& \Box_{\kappa+\lambda} \phi_\lambda^a=0\,,\qquad \lambda = \pm 1\,,
\\[5pt]
\label{man01112010-03}&& \Box_{\kappa+\lambda} \phi_\lambda=0\,,\qquad \lambda =0, \pm 2\,.
\nonumber
\eeq
We see that the gauge fixed equations are decoupled.

Modified de Donder gauge and gauge-fixed equations have leftover on-shell
gauge symmetry. Namely, modified de Donder gauge \rf{man01112010-16} and
gauge-fixed equations \rf{man01112010-02} are invariant under gauge
transformations given in \rf{man05112010-03a1} provided
the gauge transformation parameters satisfy the equations
\be
\label{man15112010-01} \Box_\kappa\xi^a= 0\,, \qquad
\Box_{\kappa+\lambda}\xi_\lambda= 0 \,,\qquad \lambda =\pm1\,. \ee

\subsection{ AdS/CFT correspondence for anomalous current
and normalizable modes of massive AdS field }\label{sec10-b}

We now ready to discuss AdS/CFT correspondence for bulk
spin-2 massive AdS field and boundary spin-2 anomalous conformal current%
\footnote{ To our knowledge AdS/CFT correspondence for bulk spin-2 massive
AdS field and boundary spin-2 anomalous conformal current has not studied in
the literature.}. To this end we use our CFT adapted approach to AdS field
dynamics and modified de Donder gauge.

First of all we note that the normalizable solution of equations of motion
\rf{man01112010-02} is given by
\beq
&& \phi^{ab}(x,z) = U_\kappa \phi_{\cur,0}^{ab}(x)\,,
\nonumber\\[5pt]
&& \phi_{-1}^a(x,z) =  - U_{\kappa-1}\phi_{\cur,-1}^a(x)\,,
\nonumber\\[5pt]
&& \phi_1^a(x,z) = U_{\kappa+1} \phi_{\cur,1}^a(x)\,,
\nonumber\\[-6pt]
&& \label{man05112010-06}
\\[-6pt]
&& \phi_{-2}(x,z) = U_{\kappa-2} \phi_{\cur,-2}(x)\,,
\nonumber\\[5pt]
&& \phi_0(x,z) = - U_\kappa\phi_{\cur,0}(x)\,,
\nonumber\\[5pt]
&& \phi_2(x,z) = U_{\kappa+2} \phi_{\cur,2}(x)\,,
\nonumber
\eeq
where $U_\nu$ is defined in \rf{man01112010-31}. From \rf{man05112010-06}, we
find the asymptotic behavior of the normalizable solution
\beq
&& \phi^{ab}(x,z) \ \ \stackrel{z\rightarrow 0}{\longrightarrow} \ \
z^{\kappa + \half} \phi_{\cur,0}^{ab}(x)\,,
\nonumber\\[5pt]
&& \phi_{-1}^a(x,z) \ \ \stackrel{z\rightarrow 0}{\longrightarrow} \ \ -
2\kappa z^{\kappa - \half} \phi_{\cur,-1}^a(x)\,,
\nonumber\\[5pt]
&& \phi_1^a(x,z) = \ \ \stackrel{z\rightarrow 0}{\longrightarrow} \ \
\frac{z^{\kappa + \frac{3}{2} } }{2(\kappa+1)} \phi_{\cur,1}^a(x)\,,
\nonumber\\[-5pt]
\label{man05112010-12} &&
\\[-5pt]
&& \phi_{-2}(x,z) = \ \ \stackrel{z\rightarrow 0}{\longrightarrow} \ \
4\kappa(\kappa+1) z^{\kappa - \frac{3}{2}} \phi_{\cur,-2}(x)\,,\qquad\quad
\nonumber\\[5pt]
&& \phi_0(x,z) = \ \ \stackrel{z\rightarrow 0}{\longrightarrow} \ \ -
z^{\kappa + \half} \phi_{\cur,0}(x)\,,
\nonumber\\[5pt]
&& \phi_2(x,z) = \ \ \stackrel{z\rightarrow 0}{\longrightarrow} \ \
\frac{z^{\kappa + \frac{5}{2}}}{4\kappa(\kappa-1)} \phi_{\cur,2}(x)\,.
\nonumber
\eeq
From \rf{man05112010-12}, we see that the fields $\phi_{\cur,0}^{ab}$,
$\phi_{\cur,\pm1}^a$, $\phi_{\cur,0}$, $\phi_{\cur,\pm 2}$ are indeed
boundary values of the normalizable solution. Moreover, in the r.h.s.
\rf{man05112010-06}, we use the notation $\phi_{\cur,0}^{ab}$,
$\phi_{\cur,\pm1}^a$, $\phi_{\cur,0}$, $\phi_{\cur,\pm 2}$ because these
boundary values turn out to be the gauge fields  entering our gauge invariant
formulation of the spin-2 anomalous conformal current in  Sec. \ref{sec05a}.
Namely, one can prove the following statements:
\\
{\bf i}) {\it Leftover on-shell} gauge transformations
\rf{man05112010-03a1} of normalizable solution
\rf{man05112010-06} lead to gauge transformations
of the anomalous conformal current \rf{man25102010-07}
\footnote{ Transformations given in \rf{man05112010-03a1} are off-shell gauge
transformations. Leftover on-shell gauge transformations are obtained from
\rf{man05112010-03a1} by using gauge transformation parameters which satisfy
equations \rf{man15112010-01}.}.
\\
{\bf ii}) For normalizable solution \rf{man05112010-06}, modified de Donder
gauge condition \rf{man01112010-16} leads to differential constraints
\rf{man24102010-40}-\rf{man24102010-42} of the anomalous conformal current.
\\
{\bf iii}) On-shell global $so(d,2)$ bulk symmetries of the normalizable
spin-2 massive modes in $AdS_{d+1}$  become global $so(d,2)$ boundary
conformal symmetries of the spin-2 anomalous conformal current.

These statements can be proved following procedure we demonstrated for the
spin-1 fields in Sec.\ref{sec09-b}. Therefore to avoid repetitions we briefly
discuss some necessary details.

{\bf Matching of bulk and boundary gauge symmetries}. To match gauge
symmetries we analyze leftover on-shell gauge symmetries which are described
by solutions of equations given in \rf{man15112010-01}. Normalizable solution
to these equations takes the form,
\beq
&& \xi^a(x,z) = U_\kappa \xi_{\cur,0}^a(x)\,,
\nonumber\\[5pt]
\label{2m26122010-01} && \xi_{-1}(x,z) = -U_{\kappa-1}\xi_{\cur,-1}(x)\,,
\\[5pt]
&& \xi_1(x,z) = U_{\kappa+1} \xi_{\cur,1}(x)\,.
\nonumber
\eeq
Plugging \rf{man05112010-06} and \rf{2m26122010-01} into bulk gauge
transformations \rf{man05112010-03a1} we make sure that the leftover on-shell
bulk gauge transformations amount to boundary gauge transformations of the
spin-2 anomalous conformal current given in \rf{man25102010-07}.

{\bf Matching of bulk de Donder gauge and boundary differential constraints}.
Plugging solution to equations for AdS fields \rf{man05112010-06} into the
modified de Donder gauge and using relations
\rf{man01112010-35},\rf{man01112010-36}, we make sure that modified de Donder
gauge \rf{man01112010-16} amounts to differential constrains
\rf{man24102010-40}-\rf{man24102010-42}.

{\bf Matching of bulk and boundary global symmetries}. Matching of bulk and
boundary Poincar\'e symmetries is obvious. Using conformal dimensions for the
spin-2 anomalous current given in \rf{man24102010-36}, solution for bulk
fields in \rf{man05112010-06}, and bulk dilatation operator
\rf{conalggenlis03ads} we make sure that dilatation bulk and boundary
symmetries also match. As before, what is non-trivial is to match $K^a$
symmetries. As in the case of the modified Lorentz gauge, the modified de
Donder gauge breaks bulk $K^a$ symmetries. In order to restore these broken
$K^a$ symmetries we add compensating gauge transformations to the generic
$K^a$ symmetries,
\be \label{man18112010-01}  K_\impr^a = K^a + \delta_{\xi^{K^a}}\,. \ee
The compensating gauge transformation parameters can as usually be found by
requiring improved transformations \rf{man18112010-01} to maintain the
modified de Donder gauge \rf{man01112010-16},
\be \label{man18112010-02} K_\impr^a C^b=0,\quad K_\impr^a C_{-1} = 0,\quad
K_\impr^a C_1=0\,.\ee
Doing so, we make sure that equations \rf{man18112010-02} amount to the
equations for the compensating gauge transformation parameters,
\beq \label{man18112010-03}
&& \Box_\kappa \xi^{bK^a} = 2 \phi^{ab} - \eta^{ab}\phi^{cc}\,,
\nonumber\\[5pt]
\label{man18112010-04} && \Box_{\kappa-1} \xi_{-1}^{K^a} = 2 \phi_{-1}^a\,,
\\[5pt]
\label{man18112010-05} && \Box_{\kappa+1} \xi_1^{K^a} = 2 \phi_1^a\,.
\nonumber
\eeq
Using \rf{man01112010-37a1} and \rf{man05112010-06}, we find solution for the
compensating gauge transformation parameters,
\beq \label{man18112010-06}
&& \xi^{bK^a}(x,z) = zU_{\kappa+1} (\phi_{\cur,0}^{ab}(x) - \half
\eta^{ab}\phi_{\cur,0}^{cc}(x))\,,\qquad\quad
\nonumber\\[5pt]
\label{man18112010-07}  && \xi_{-1}^{K^a}(x,z) =  - z
U_\kappa\phi_{\cur,-1}^a(x)\,,
\\[5pt]
&& \xi_1^{K^a}(x,z) = zU_{\kappa+2} \phi_{\cur,1}^a(x)\,,
\nonumber
\eeq
where operator $U_\nu$ is given in \rf{man01112010-31}. Plugging
\rf{man05112010-06} and \rf{man18112010-07} in \rf{man18112010-01}, we make
sure that the improved bulk $K_\impr^a$ symmetries of the spin-2 massive AdS
field amount to $K^a$ symmetries of the spin-2 anomalous conformal current
given in \rf{conalggenlis04} and \rf{man25102010-12a1}.

\subsection{ AdS/CFT correspondence for anomalous shadow field
and non-normalizable mode of massive AdS field.}

We proceed to discussion of AdS/CFT correspondence for bulk spin-2 massive
AdS field and boundary spin-2 anomalous shadow field.

{\bf Matching of effective action and boundary two-point vertex}. In order to
find $S_\eff$ we should solve equations of motion with the Dirichlet problem
corresponding to the boundary anomalous shadow field and plug the solution
into action. Using equations of motion \rf{man14112010-01} in bulk action
\rf{09072009-02} with Lagrangian \rf{man29102010-11}, we obtain boundary
effective action \rf{man27102010-05} with $\LL_\eff$ given by
\beq \label{man01112010-27}
\LL_\eff & = & \frac{1}{4} \phi^{ab} \TT_{\kappa-\half}\phi^{ab} - \frac{1}{8}
\phi^{aa} \TT_{\kappa -\half } \phi^{bb}
\nonumber\\[5pt]
& + &  \half \sum_{\lambda=\pm1} \phi_\lambda^a \TT_{\kappa+\lambda -\half } \phi_\lambda^a
\nonumber\\[5pt]
& + & \half \sum_{\lambda=0,\pm2} \phi_\lambda \TT_{\kappa+\lambda -\half } \phi_\lambda\,,
\nonumber\\[5pt]
& - & \half (r_z^{00}\phi_{-1}^a + r_\zeta^{00}\phi_1^a) C^a
\nonumber\\[5pt]
& + & (\frac{r_z^{00}}{4}\phi^{aa} -\frac{r_z^{01}}{\sqrt{2}}\phi_{-2} -
\frac{r_\zeta^{01}}{2}\phi_0)C_{-1}
\nonumber\\[5pt]
& + & (\frac{r_\zeta^{00}}{4}\phi^{aa} -\frac{r_z^{10}}{2}\phi_0 -
\frac{r_\zeta^{10}}{\sqrt{2}}\phi_2)C_1\,.
\eeq
We have demonstrated that the use of the modified de Donder gauge
considerable simplifies the equations of motion. Now using modified de Donder
gauge \rf{man01112010-16} in \rf{man01112010-27}, we obtain
\beq \label{man01112010-23}
\LL_\eff\Bigr|_{{\,C^a= 0 \atop C_{\pm1}=0}} & = & \frac{1}{4} \phi^{ab}
\TT_{\kappa-\half}\phi^{ab} - \frac{1}{8} \phi^{aa} \TT_{\kappa -\half }
\phi^{bb}
\nonumber\\[5pt]
& + &  \half \sum_{\lambda=\pm1} \phi_\lambda^a \TT_{\kappa -\half +\lambda  } \phi_\lambda^a
\nonumber\\[5pt]
& + & \half \sum_{\lambda=0,\pm2} \phi_\lambda \TT_{\kappa-\half +\lambda  } \phi_\lambda\,,
\eeq
i.e. we see that $\LL_\eff$ is also considerably simplified. To find $S_\eff$
we should solve gauge-fixed equations of motion \rf{man01112010-02} with the
Dirichlet problem corresponding to the boundary anomalous shadow field and
plug the solution into $\LL_\eff$. To this end we discuss solution of
equations of motion \rf{man01112010-02}.

As before our equations of motion take decoupled form and similar to the
equations of motion for the massive scalar AdS field. Therefore we can
apply the procedure described in Sec. \ref{secAdS/CFT}. Doing so, we obtain
solution of equation \rf{man01112010-02} with the Dirichlet problem
corresponding to the spin-2 anomalous shadow field,
\beq
&& \label{man01112010-07} \hspace{-0.7cm} \phi^{ab}(x,z)  = \sigma_{2,0}\!
\int\! d^dy\, G_{\kappa}(x-y,z) \phi_{\sh,0}^{ab}(y)\,, \qquad
\\[5pt]
&& \label{man01112010-08} \hspace{-0.7cm}  \phi_\lambda^a(x,z)  =
\sigma_{1,\lambda}\! \int\! d^dy\, G_{\kappa+\lambda}(x-y,z)
\phi_{\sh,-\lambda}^a(y)\,,
\nonumber\\
&& \hspace{0.7cm} \lambda = \pm 1\,,
\\ [5pt]
&& \label{man01112010-09} \hspace{-0.7cm} \phi_\lambda(x,z)  =
\sigma_{0,\lambda}\! \int\! d^dy\, G_{\kappa+\lambda}(x-y,z)
\phi_{\sh,-\lambda}(y)\,,
\nonumber\\
&& \hspace{0.7cm} \lambda = 0,\pm 2\,,
\\[5pt]
\label{man01112010-10} && \hspace{-0.3cm}\sigma_{2,0} = 1\,,
\\[5pt]
&& \hspace{-0.3cm} \sigma_{1,-1} = -\frac{1}{2(\kappa-1)}\,,
\hspace{1.1cm} \sigma_{1,1} = 2\kappa\,,
\nonumber\\[5pt]
\label{man01112010-11} && \hspace{-0.3cm} \sigma_{0,-2} =
\frac{1}{4(\kappa-1)(\kappa-2)}\,,
\\[5pt]
&& \hspace{-0.3cm} \sigma_{0,0} = -1 \,,
\hspace{1.5cm} \sigma_{0,2} = 4\kappa(\kappa+1)\,,\qquad\quad
\nonumber
\eeq
where the Green function $G_\nu$ is given in \rf{10072009-12}, while $\kappa$
is defined in \rf{man24102010-39}. Choice of normalization factor
$\sigma_{2,0}$ \rf{man01112010-10} is a matter of convention. The remaining
normalization factors given in \rf{man01112010-11} are uniquely determined by
requiring that modified de Donder gauge \rf{man01112010-16} be amount to the
differential constraints for the spin-2 anomalous shadow field.

Using asymptotic behavior of the Green function given in
\rf{10072009-14}, we find the asymptotic behavior of our solution
\beq
&& \phi^{ab}(x,z) \,\,\, \stackrel{z\rightarrow 0 }{\longrightarrow} \,\,\,
z^{-\kappa+\half}\phi_{\sh,0}^{ab}(x)\,,
\nonumber\\[5pt]
&& \phi_{-1}^a(x,z) \,\,\, \stackrel{z\rightarrow 0 }{\longrightarrow}\,\,\,
-\frac{z^{-\kappa+\frac{3}{2}}}{2(\kappa-1)}\phi_{\sh,1}^a(x)\,,
\nonumber\\[5pt]
&& \phi_1^a(x,z) \,\,\, \stackrel{z\rightarrow 0 }{\longrightarrow}\,\,\,
2\kappa z^{-\kappa-\half}\phi_{\sh,-1}^a(x)\,,
\nonumber\\[-6pt]
\label{man01112010-17} &&
\\[-6pt]
&& \phi_{-2}(x,z)\,\,\, \stackrel{z\rightarrow 0 }{\longrightarrow}\,\,\,
\frac{z^{-\kappa+\frac{5}{2}}}{4(\kappa-1)(\kappa-2)}\phi_{\sh,2}(x)\,,
\nonumber\\[5pt]
&& \phi_0(x,z) \,\,\, \stackrel{z\rightarrow 0 }{\longrightarrow}\,\,\, -
z^{-\kappa+\half} \phi_{\sh,0}(x)\,,
\nonumber\\[5pt]
&& \phi_2(x,z) \,\,\, \stackrel{z\rightarrow 0 }{\longrightarrow}\,\,\,
4\kappa(\kappa+1)z^{-\kappa-\frac{3}{2}}\phi_{\sh,-2}(x)\,,\qquad
\nonumber
\eeq
which tells us that solution \rf{man01112010-07}-\rf{man01112010-09} has
indeed asymptotic behavior corresponding to the anomalous shadow field.

Finally, to obtain the effective action we plug solution of the Dirichlet
problem for AdS fields, \rf{man01112010-07}-\rf{man01112010-09} into
\rf{man27102010-05}, \rf{man01112010-23}. Using general formula given in
\rf{man02-21072009-15}, we obtain
\be \label{man08112010-03}
-S_\eff = 2\kappa c_\kappa \Gamma \,,
\ee
where $\kappa$ and $c_\kappa$ are defined in \rf{man24102010-39} and
\rf{10072009-13} respectively and $\Gamma$ is gauge invariant two-point
vertex of the spin-2 anomalous shadow field given in
\rf{man05112010-18},\rf{man07112010-04}.

Thus, {\it using the modified de Donder gauge for the spin-2 massive AdS
field and computing the bulk action on solution of equations of motion with
the Dirichlet problem corresponding to the boundary anomalous shadow field we
obtain the gauge invariant two-point vertex of the spin-2 anomalous shadow
field}.

Using \rf{man07112010-05}, we can represent our result \rf{man08112010-03} in
the Stueckelberg gauge frame
\be \label{man08112010-04} - S_\eff =
\frac{\kappa(2\kappa+d+2)}{2(2\kappa+d-2)} c_\kappa \Gamma^{{\rm stand}}
\,,\ee
while, in the light-cone gauge frame, our result is represented as
\be \label{2m26122010-03}
-S_\eff = 2\kappa c_\kappa \Gamma^{({\rm l.c.})} \,,
\ee
where $\Gamma^{({\rm l.c.})}$ is given in \rf{man07112010-15}. Relation
\rf{man08112010-04} with the normalization factor in front of $\Gamma^{{\rm
stand}}$ as in \rf{man08112010-04} was obtained in
Ref.\cite{Polishchuk:1999nh}%
\footnote{ Computation of $S_\eff$ for spin-2 massless field may be found in
Refs.\cite{Liu:1998bu,Arutyunov:1998ve,Mueck:1998ug}. In the framework of our
approach, $S_\eff$ was studied in Ref.\cite{Metsaev:2009hp}.}.
Note that we have obtained more general relation given in
\rf{man08112010-03}, while relation \rf{man08112010-04} is obtained from
\rf{man08112010-03} by using the Stueckelberg gauge frame. It is our general
relation \rf{man08112010-03} that provides possibility for the derivation of
all other relations like the ones in \rf{man08112010-04} and
\rf{2m26122010-03} just by choosing appropriate gauge conditions. Note that
transformation of relation \rf{man08112010-04} to the one in
\rf{2m26122010-03} requires cumbersome computations because the Stueckelberg
gauge frame removes the vector and scalar field entering the light-cone gauge
frame (see Secs.\ref{sec006-b} and \ref{sec006-c}).

{\bf Matching of bulk and boundary gauge symmetries}. Modified de Donder
gauge \rf{man01112010-16} and gauge-fixed equations \rf{man01112010-02} are
invariant under gauge transformations given in \rf{man05112010-03a1} provided
the gauge transformation parameters satisfy equations \rf{man15112010-01}.
The non-normalizable solution to equations \rf{man15112010-01} is given by
\beq
\label{man15112010-02} && \xi^a(x,z)  =  \int d^dy\,
G_\kappa(x-y,z)\xi_{\sh,0}(y)\,,
\nonumber\\[-6pt]
&&
\\[-6pt]
&& \xi_\lambda(x,z)  = \sigma_{1,\lambda}\!\!\int d^dy\,
G_{\kappa+\lambda}(x-y,z)\xi_{\sh,-\lambda}(y)\,,\qquad\quad
\nonumber
\eeq
$\lambda=\pm 1$, where $\sigma_{1,\pm 1}$ are given in \rf{man01112010-11}.
Plugging \rf{man15112010-02} and \rf{man01112010-07}-\rf{man01112010-09} in
\rf{man05112010-03a1} we make sure the on-shell leftover gauge symmetries of
solution of the Dirichlet problem for spin-2 massive AdS field amount to the
gauge symmetries of the spin-2 anomalous shadow field \rf{man29102010-03}.

{\bf Matching of bulk and boundary global symmetries}. The matching can be
demonstrated by following the procedure we used for the spin-2 anomalous
current in Sec.\ref{sec10-b}. Therefore to avoid repetitions we briefly
discuss some necessary details. Matching of bulk and boundary Poinca\'e
symmetries is obvious. Using conformal dimensions for the spin-2 anomalous
shadow given in \rf{man25102010-14n}, solution for bulk fields in
\rf{man01112010-07}-\rf{man01112010-09}, and bulk dilatation operator
\rf{conalggenlis03ads}, we make sure that dilatation bulk and boundary
symmetries also match. In order to match $K^a$ symmetries we consider
improved $K_\impr^a$ transformations \rf{man18112010-01} with gauge
transformation parameters that satisfy equations \rf{man18112010-03}. Using
\rf{2m26122010-04}, we see that solution to equations \rf{man18112010-03}
with right hand sides as in \rf{man01112010-07}, \rf{man01112010-08} is given
by
\beq
&& \hspace{-1cm} \xi^{bK^a}(x,z) = z\sigma_{2,0}^\xi\! \int\! d^dy\,
G_{\kappa-1}(x-y,z)
\nonumber\\[3pt]
&&  \hspace{1cm} \times \,\, (\phi_{\sh,0}^{ab}(y) - \half
\eta^{ab}\phi_{\sh,0}^{cc})\,,\qquad\quad
\nonumber\\[5pt]
&& \hspace{-1cm} \xi_{-1}^{K^a}(x,z) =  z \sigma_{1,-1}^\xi\! \int\!\! d^dy\,
G_{\kappa-2}(x-y,z) \phi_{\sh,1}^a(y)\,,\qquad
\\[5pt]
&& \hspace{-1cm} \xi_1^{K^a}(x,z) = z \sigma_{1,1}^\xi\! \int\! d^dy\,
G_{\kappa}(x-y,z) \phi_{\sh,-1}^a(y)\,,
\nonumber\\[3pt]
&& \hspace{-0.7cm} \sigma_{2,0}^\xi \equiv  -\frac{1}{2(\kappa-1)}\,,
\\[3pt]
&& \hspace{-0.7cm} \sigma_{1,-1}^\xi \equiv \frac{1}{4(\kappa-1)(\kappa-2)}\,,
\qquad \sigma_{1,1}^\xi \equiv -1\,,
\eeq
where the Green function is given in \rf{10072009-12}. Using these
compensating gauge transformation parameters in improved bulk $K_\impr^a$
symmetries \rf{man18112010-01} we make sure that these $K_\impr^a$ symmetries
amount to $K^a$ symmetries of spin-2 anomalous shadow field given in
\rf{conalggenlis04} and \rf{man11112010-04}.

To summarize, it is {\it the matching of the bulk on-shell leftover gauge
symmetries of the solution to Dirichlet problem and bulk global symmetries
and the respective boundary gauge symmetries of the anomalous shadow field
and boundary global symmetries that explains why the effective action
coincides with the gauge invariant two-point vertex for the boundary
anomalous shadow field (see \rf{man08112010-03})}.

Comparing our results for spin-1 and spin-2 fields given in
\rf{man08112010-01} and \rf{man08112010-03} respectively, we see that our
approach gives uniform description of the interrelation between the effective
action of massive AdS fields and two-point gauge invariant vertex of shadow
fields. Note however that value of $\kappa$ for spin-1 field
\rf{man24102010-02n} should not be confused with the one for spin-2 field
\rf{man24102010-39}. For the case of arbitrary spin-$s$ field, the $\kappa$
was found in Refs.\cite{Metsaev:2003cu,Metsaev:2009hp},
\be \label{man19112010-03} \kappa = \sqrt{m^2 + \bigl(s+
\frac{d-4}{2}\bigr)^2}\,.\ee
All that is required to generalize relation \rf{man08112010-03} to arbitrary
spin-$s$ fields is to plug $\kappa$ \rf{man19112010-03} in
\rf{man08112010-03}. Detailed study of arbitrary spin fields will be given in
forthcoming publication.

\section{Conclusions}\label{conl-sec-01}

In this paper, we extend the gauge invariant Stueckelberg approach to CFT
initiated in Refs.\cite{Metsaev:2008fs,Metsaev:2009ym} to the study of
anomalous conformal currents and shadow fields. In the framework of AdS/CFT
correspondence the anomalous conformal currents and shadow fields are related
with massive fields of AdS string theory. It is well known that all Lorentz
covariant approaches to string field theory involve large amount of
Stueckelberg fields and the corresponding gauge symmetries (see e.g.
\cite{Siegel:1985tw}). Because our approach to anomalous conformal currents
and shadow fields also involves Stueckelberg fields we believe that our
approach will be helpful to understand string/gauge theory duality better.
Note also that we obtain gauge invariant vertex for anomalous shadow fields
which provides quick and easy access to light-cone gauge vertex. In the
framework of AdS/CFT correspondence this vertex is related to AdS field
action evaluated on solution of the Dirichlet problem. Because on expects
that quantization of AdS superstring is straightforward only in light-cone
gauge we believe that our light-cone gauge vertex will also be helpful in
various studies of AdS/CFT duality. The results obtained should have a number
of the following interesting applications and generalizations.

(i) In this paper, we considered the gauge invariant approach for spin-1 and
spin-2 anomalous conformal currents and shadow fields. It would be
interesting to generalize our approach to the case of arbitrary spin
anomalous conformal currents and shadow fields.

(ii) In this paper we studied the two-point gauge invariant vertex of
anomalous shadow fields. Generalization of our approach to the case of
3-point and 4-point gauge invariant vertices will give us the possibility to
the study of various applications of our approach along the lines of
Refs.\cite{Roiban:2010fe,Liu:1998ty,Leonhardt:2002ta}

(iii) Because our modified de Donder gauge leads to considerably simplified
analysis of AdS field dynamics we believe that this gauge might also be
useful for better understanding of various aspects of AdS/QCD correspondence
which are discussed e.g. in Refs.\cite{Andreev:2002aw,Brodsky:2008pg}.

iv) BRST approach is one of powerful approaches to analysis of various
aspects of relativistic dynamics (see e.g.
Refs.\cite{Siegel:1999ew}-\cite{Alkalaev:2008gi}. We think that extension of
this approach to the case anomalous conformal currents and shadow fields
should be relatively straightforward.

v) In the last years, there were interesting developments in studying the
mixed symmetry fields \cite{Alkalaev:2003qv}-\cite{Skvortsov:2008vs}. It
would be interesting to apply methods developed in these references to
studying anomalous conformal currents and shadow fields. There are other
various interesting approaches in the literature which could be used to
discuss gauge invariant formulation of anomalous conformal currents and
shadow fields. This is to say that various recently developed interesting
formulations of field dynamics in terms of unconstrained fields in flat space
may be found in Refs.\cite{Francia:2002aa}-\cite{Campoleoni:2008jq} .

\begin{acknowledgments}
This work was supported by the RFBR Grant No.08-02-01118, by the Dynasty
Foundation and by the Alexander von Humboldt Foundation Grant PHYS0167.
\end{acknowledgments}

\appendix
\section{ Derivation of CFT adapted Lagrangian for spin-1 massive field
in $AdS_{d+1}$ } \label{app-01}

In this Appendix, we explain some details of the derivation of the CFT
adapted gauge invariant Lagrangian for spin-1 massive field given in
\rf{man27102010-01}. Presentation in this Appendix is given by using
Lorentzian signature. Euclidean signature Lagrangian in Sec.\ref{sec09-a}, is
obtained from the Lorentzian signature Lagrangian by simple substitution
$\LL\rightarrow -\LL$.

{\bf Spin-1 massive field}. We use field $\Phi^A$ carrying flat Lorentz
algebra $so(d,1)$ vector indices $A,B=0,1,\ldots, d-1,d$. The field $\Phi^A$
is related with field carrying the base manifold indices $\Phi^\mu$,
$\mu=0,1,\ldots,d$, in standard way $\Phi^A= e_\mu^A\Phi^\mu$, where
$e_\mu^A$ is vielbein of $AdS_{d+1}$ space. For the Poincar\'e
parametrization of $AdS_{d+1}$ space \rf{lineelem01}, vielbein $e^A=e^A_\mu
dx^\mu$ and Lorentz connection, $de^A+\omega^{AB}\wedge e^B=0$, are given by
\be\label{eomcho01} e_\mu^A=\frac{1}{z}\delta^A_\mu\,,\qquad
\omega^{AB}_\mu=\frac{1}{z}(\delta^A_z\delta^B_\mu
-\delta^B_z\delta^A_\mu)\,, \ee
where $\delta_\mu^A$ is Kronecker delta symbol. We use a covariant derivative
with the flat indices $\DD^A$,
\be \DD_A \equiv e_A^\mu \DD_\mu\,,\qquad \DD^A = \eta^{AB}\DD_B\,,\ee
where $e_A^\mu$ is inverse of AdS vielbein, $e_\mu^A e_B^\mu = \delta_B^A$
and $\eta^{AB}$ is flat metric tensor. With choice made in \rf{eomcho01}, the
covariant derivative takes the form
\be
\DD^A \Phi^B = \hat{\partial}^A \Phi^B + \delta_z^B \Phi^A -
\eta^{AB}\Phi^z\,,\quad \ \hat{\partial}^A \equiv z\partial^A\,,
\ee
where we adopt the following conventions for the derivatives and coordinates:
$\partial^A=\eta^{AB}\partial_B$, $\partial_A=\partial/\partial x^A $, $x^A
\equiv \delta_\mu^A x^\mu$, $x^A=x^a,x^d$ with the identification $x^d\equiv
z$.

In arbitrary parametrization of AdS, Lagrangian of the spin-1 massive field
is given in \rf{man11112010-11}. We now use the Poincar\'e parametrization of
AdS and introduce the following quantity:
\be
\Cbf \label{man13112010-14} \equiv \DD^C\Phi^C + m \Phi + 2\Phi^z  \,.
\ee
We note that it is the relation $\Cbf=0$ that defines the modified Lorentz
gauge. Using the relations (up to total derivative)
\beq
\label{man02-13072009-05} e\Phi^A \DD^2 \Phi^A & = & e\Bigl(\Phi^A
(\Box_{_{0\,AdS}} -1) \Phi^A
\nonumber\\[5pt]
&  + & 4 \Phi^z \Cbf + (d-7) \Phi^z \Phi^z - 4m \Phi\Phi^z\Bigr)\,,\qquad
\\[5pt]
\label{man02-13072009-05a} e\Phi \DD^2 \Phi & = & e\Phi \Box_{_{0\,AdS}} \Phi
\\[5pt]
\label{man02-13072009-06} C_\st^2 & = &  \Cbf^2 - 4 \Phi^z \Cbf + 4 \Phi^z
\Phi^z\,,
\\[5pt]
\label{man02-13072009-07} \Box_{_{0\,AdS}} & \equiv & z^2(\Box+\partial_z^2)
+(1-d)z\partial_z \,,
\eeq
$e\equiv \det e_\mu^A$, we represent Lagrangian \rf{man11112010-11} and
$\Cbf$ \rf{man13112010-14} as
\beq
\label{man02-13072009-08} e^{-1} \LL & = &  \half \Phi^A (\Box_{_{0\,AdS}}
-m^2 + d-1 ) \Phi^A
\nonumber\\[5pt]
& + & \half \Phi (\Box_{_{0\,AdS}} - m^2) \Phi
\nonumber\\[5pt]
& + & \frac{d-3}{2}\Phi^z\Phi^z  - 2m \Phi \Phi^z + \half \Cbf^2 \,,
\\[5pt]
\label{man02-13072009-09} \Cbf  & = &   \hat\partial^A \Phi^A + (2-d)\Phi^z +
m\Phi \,.\eeq
Using canonically normalized fields $\widetilde\Phi^A$, $\widetilde\Phi$ and
$C$ defined by
\be \Phi^A = z^{\frac{d-1}{2}}\widetilde\Phi^A\,,\qquad \Phi =
z^{\frac{d-1}{2}}\widetilde\Phi\,,\qquad \Cbf = z^{\frac{d+1}{2}} C \,, \ee
we obtain
\beq  \label{man13112010-15}
\LL & = &  \half \widetilde\Phi^A \Bigl( \Box +\partial_z^2  - \frac{1}{z^2}
(m^2 + \frac{d^2-1}{4}+ 1-d)\Bigr) \widetilde\Phi^A
\nonumber\\[5pt]
& + & \half \widetilde\Phi \Bigl(\Box +\partial_z^2  - \frac{1}{z^2} (m^2 +
\frac{d^2 -1}{4})\Bigr) \widetilde\Phi
\nonumber\\[5pt]
& + & \frac{d-3}{2z^2} \widetilde\Phi^z \widetilde\Phi^z -
\frac{2m}{z^2}\widetilde\Phi^z \widetilde\Phi + \half C^2 \,,
\\[5pt]
\label{man13112010-16} C  & = & \partial^A \widetilde\Phi^A + \frac{3-d}{2z}
\widetilde\Phi^z + \frac{m}{z} \widetilde\Phi\,.
\eeq
In terms of $so(d-1,1)$ tensorial components of the field $\widetilde\Phi^A$
given by $\widetilde\Phi^a$, $\widetilde\Phi^z$, Lagrangian
\rf{man13112010-15} and $C$ \rf{man13112010-16} take the form
\beq  \label{man13112010-17}
\LL & = & \LL_1 + \LL_0  + \half C^2\,,
\\[5pt]
\label{man13112010-17a1}
&& \LL_1 =  \half \widetilde\Phi^a \hat{K}_0 \widetilde\Phi^a
\\[5pt]
\label{man13112010-17a2}&& \LL_0 = \half
\widetilde\Phi^z \hat{K}_{3-d} \widetilde\Phi^z + \half \widetilde\Phi
\hat{K}_{d-1} \widetilde\Phi - \frac{2m}{z^2} \widetilde\Phi^z \widetilde\Phi\,,\qquad\quad
\\[5pt]
\label{man13112010-17a3} && C  = \partial^a \widetilde\Phi^a +
\TT_{\frac{3-d}{2}}\widetilde\Phi^z + \frac{m}{z} \widetilde\Phi\,,
\\[5pt]
\label{khatdef}
&& \hspace{1cm} \hat{K}_\omega = \Box +\partial_z^2 - \frac{1}{z^2}(\kappa^2  - \frac{1}{4}
+\omega)\,,
\eeq
where $\kappa$ and $\TT_\nu$ are defined in \rf{man24102010-02n} and
\rf{09072009-04} respectively. In terms of fields \rf{man11112010-14} defined
by
\beq
&& \widetilde\Phi^a = \phi^a\,,
\nonumber\\[5pt]
\label{man13112010-18} && \widetilde\Phi^z = r_z^{00}\phi_{-1} + r_\zeta^{00}
\phi_1\,,
\\[5pt]
&& \widetilde\Phi = - r_\zeta^{00}\phi_{-1} + r_z^{00}\phi_1\,,
\nonumber
\eeq
where $r_z^{00}$, $r_\zeta^{00}$ are defined in \rf{man24102010-04}, we
represent $\LL_1$ \rf{man13112010-17a1} and $\LL_0$ \rf{man13112010-17a2} as
\beq
&& \LL_1 = \half \phi^a
\Box_\kappa \phi^a\,,
\nonumber\\[-6pt]
\label{man13112010-20a1} &&
\\[-6pt]
&& \LL_0 = \half \sum_{\lambda= \pm 1} \phi_\lambda
\Box_{\kappa+\lambda} \phi_\lambda\,,
\nonumber
\eeq
while $C$ \rf{man13112010-17a3} takes desired form given in \rf{11082009-03}.
Noticing the relation
\be \label{man13112010-57}  \TT_{\nu-\half}^\dagger\TT_{\nu-\half} =
-\partial_z^2 + \frac{1}{z^2}(\nu^2-\frac{1}{4})\,, \ee
and taking into account expressions for $\Box_\nu$ \rf{09072009-09} and
$\LL_1$, $\LL_0$ \rf{man13112010-20a1},  we see that Lagrangian
\rf{man13112010-17} takes the form of the CFT adapted gauge invariant
Lagrangian \rf{man27102010-01}.

Lagrangian \rf{man11112010-11} is invariant under gauge transformations
\rf{man11112010-10a}. Making the rescaling $\Xi = z^{(d-3)/2}\xi$, we check
that these gauge transformations lead to the ones given in
\rf{11082009-04}-\rf{11082009-06}.

\section{ Derivation of CFT adapted Lagrangian for spin-2 massive field
in $AdS_{d+1}$ }\label{app-02}

We present details of the derivation of the CFT adapted gauge invariant
Lagrangian and the respective gauge transformations of spin-2 massive field
given in \rf{man29102010-11} and \rf{man05112010-03a1}.

In arbitrary parametrization of AdS, Lagrangian for the spin-2 massive field
is given in \rf{man11112010-21}. We now use the Poincar\'e parametrization of
AdS and introduce the following quantities
\beq
&& \Cbf^A \equiv C_\st^A  + 2 \Phi^{zA} - \delta_z^A \Phi^{BB} \,,
\nonumber\\[-6pt]
\label{man13112010-21} &&
\\[-6pt]
&& \Cbf \equiv C_\st  + 2 \Phi^z\,.
\nonumber
\eeq
We note that it is the relations $\Cbf^A=0$, $\Cbf=0$ that define the modified
de Donder gauge. Using the relations (up to total derivative)
\beq \label{man13112010-23}
&& \frac{1}{4}e \Phi^{AB}\DD^2 \Phi^{AB}  = e\Bigl( \frac{1}{4} \Phi^{AB}
(\Box_{_{0\,AdS}} -2)\Phi^{AB}
\nonumber\\[5pt]
&& + \frac{d-5}{2} \Phi^{zA}\Phi^{zA} + 2\Phi^{zz}\Phi^{AA}
-\frac{d}{4}\Phi^{AA}\Phi^{BB}
\nonumber\\[5pt]
&& +  2\Phi^{zA} \Cbf^A - \Phi^{AA} \Cbf^z -2m\Phi^{zA}\Phi^A +
m\Phi^{AA}\Phi^z\Bigr),\qquad
\\[10pt]
&& \half C_\st^A C_\st^A  =   \half \Cbf^A \Cbf^A - 2\Phi^{zA} \Cbf^A +
\Phi^{AA} \Cbf^z
\nonumber\\[5pt]
&& + 2\Phi^{zA}\Phi^{zA} - 2\Phi^{zz}\Phi^{AA} + \half \Phi^{AA}\Phi^{BB}\,,
\eeq
\beq \label{man13112010-24}
&& e\Phi^A \DD^2 \Phi^A  = e\Bigl(\Phi^A (\Box_{_{0\,AdS}} -1) \Phi^A
\nonumber\\[5pt]
&& + 4 \Phi^z \Cbf + (d-7) \Phi^z \Phi^z - 2m \Phi^{AA} \Phi^z  - 4f \Phi
\Phi^z\Bigr),\qquad
\\[5pt]
&& C_\st^2 =  \Cbf^2 - 4 \Phi^z \Cbf + 4 \Phi^z \Phi^z\,,
\eeq
where $\Box_{_{0\,AdS}}$ is given in \rf{man02-13072009-07}, we represent
Lagrangian \rf{man11112010-21} and $\Cbf^A$, $\Cbf$
\rf{man13112010-21} as
\beq \label{man13112010-25}
e^{-1} \LL & = & \frac{1}{4}\Phi^{AB} (\Box_{_{0\,AdS}} - m^2) \Phi^{AB}
\nonumber\\[5pt]
& - & \frac{1}{8}\Phi^{AA} (\Box_{_{0\,AdS}} - m^2) \Phi^{BB}
\nonumber\\[5pt]
& + &  \frac{d-1}{2}\Phi^{zA}\Phi^{zA}  - 2m \Phi^{zA}\Phi^A
\nonumber\\[5pt]
& +  & \half \Phi^A (\Box_{_{0\,AdS}} - m^2 -d -1 ) \Phi^A
\nonumber\\[5pt]
& + & \frac{d-3}{2}\Phi^z\Phi^z -2f \Phi\Phi^z
\nonumber\\[5pt]
& + &  \half \Phi (\Box_{_{0\,AdS}} - m^2 -2d) \Phi
\nonumber\\[5pt]
& + & \half \Cbf^A \Cbf^A + \half \Cbf\Cbf\,,
\\[5pt]
\Cbf^A & = & \hat\partial^B \Phi^{AB} - \half \hat\partial^A \Phi^{BB} +
(1-d) \Phi^{zA} + m\Phi^A\,,\qquad
\nonumber\\[-5pt]
\label{man13112010-26} &&
\\[-5pt]
\Cbf & = & \hat\partial^A \Phi^A  + (2-d) \Phi^z + \frac{m}{2} \Phi^{AA} + f
\Phi \,.
\nonumber
\eeq
Using canonically normalized fields and quantities $\tilde\Cbf^A$,
$\tilde\Cbf$,
\beq
&& \hspace{-1cm} \Phi^{AB} = z^{\frac{d-1}{2}}\tilde\Phi^{AB}\,, \quad \Phi^A
= z^{\frac{d-1}{2}}\tilde\Phi^A\,, \quad \Phi =
z^{\frac{d-1}{2}}\tilde\Phi\,,
\nonumber\\[-6pt]
\label{man13112010-28} &&
\\[-6pt]
&& \hspace{-1cm} \Cbf^A = z^{\frac{d+1}{2}}\tilde\Cbf^A\,, \qquad \Cbf =
z^{\frac{d+1}{2}}\tilde\Cbf\,,
\nonumber
\eeq
we obtain
\beq \label{man13112010-30}
\LL & = &  \frac{1}{4} \tilde\Phi^{AB} \hat{K}_0 \tilde\Phi^{AB} -\frac{1}{8}
\tilde\Phi^{AA} \hat{K}_0 \tilde\Phi^{BB}
\nonumber\\[5pt]
& + & \half \tilde\Phi^A \hat{K}_{d +1} \tilde\Phi^A + \half \tilde\Phi
\hat{K}_{2d} \tilde\Phi
\nonumber\\[5pt]
& + & \frac{d-1}{2z^2} \tilde\Phi^{zA}\tilde\Phi^{zA} - \frac{2m}{z^2}
\tilde\Phi^{zA} \tilde\Phi^A
+ \frac{d-3}{2z^2} \tilde\Phi^z \tilde\Phi^z
\nonumber\\[5pt]
& - & \frac{2 f}{z^2} \tilde\Phi^z \tilde\Phi + \half \tilde\Cbf^A
\tilde\Cbf^A + \half \tilde\Cbf\tilde\Cbf\,,\qquad
\eeq
\beq
\tilde\Cbf^a & = & \partial^b \tilde\Phi^{ab} - \half \partial^a
\tilde\Phi^{BB} + \TT_{-\frac{d-1}{2}} \tilde\Phi^{za} + \frac{m}{z}
\tilde\Phi^a \,,
\nonumber\\[5pt]
\label{man13112010-31} \tilde\Cbf^z & = & \partial^a \tilde\Phi^{za} - \half
\TT_{\frac{d-1}{2}} \tilde\Phi^{BB} + \TT_{- \frac{d-1}{2z}}\tilde\Phi^{zz} +
\frac{m}{z} \tilde\Phi^z\,,\qquad
\\[5pt]
\tilde\Cbf & = & \partial^a \tilde\Phi^a + \TT_{-\frac{d-3}{2}} \tilde\Phi^z
+ \frac{m}{2z} \tilde\Phi^{AA}  + \frac{f}{z}\tilde\Phi \,,
\nonumber
\eeq
where $\kappa$ and $\hat{K}_\omega$ are defined in \rf{man24102010-39} and
\rf{khatdef} respectively. In terms of new fields defined by the relations
\beq
&& \phi^{ab} = \tilde\Phi^{ab} + \frac{1}{d-2}\eta^{ab}\tilde\Phi^{zz}\,,
\nonumber\\[5pt]
\label{man13112010-34} && \phi^{za} = \tilde{\Phi}^{za}\,,\qquad\quad
\phi^a = \tilde{\Phi}^a\,,
\\[5pt]
&& \phi^{zz} = \frac{u}{2}\, \tilde\Phi^{zz}\,, \qquad \phi^z = \tilde{\Phi}^z\,, \qquad
\phi = \tilde{\Phi}\,,\qquad
\nonumber
\eeq
Lagrangian $\LL$ \rf{man13112010-30} and $\tilde\Cbf^A$, $\tilde\Cbf$
\rf{man13112010-31} take the form
\be \label{man13112010-37} \LL = \LL_2 + \LL_1 + \LL_0 + \half \tilde\Cbf^A
\tilde\Cbf^A + \half \tilde\Cbf\tilde\Cbf\,, \ee
\beq \label{man13112010-38}
\LL_2 & = &  \frac{1}{4} \phi^{ab} \hat{K}_0 \phi^{ab} -\frac{1}{8} \phi^{aa}
\hat{K}_0 \phi^{bb}\,,\qquad\quad
\\[5pt]
\label{man13112010-39} \LL_1 & = &  \half \phi^{za} \hat{K}_{1-d} \phi^{za}
+  \half \phi^a \hat{K}_{1+d} \phi^a - \frac{2m}{z^2} \phi^{za}
\phi^a\,,\qquad
\\[5pt]
\label{man13112010-40} \LL_0 & = & \half \phi^{zz} \hat{K}_{4-2d} \phi^{zz}
 +  \half \phi^z \hat{K}_4 \phi^z + \half \phi \hat{K}_{2d} \phi
\nonumber\\[5pt]
& - & \frac{2g}{z^2} \phi^{zz} \phi^z - \frac{2f}{z^2} \phi^z \phi \,, \qquad
\eeq

\beq
\tilde\Cbf^a & = & \partial^b \phi^{ab} -\half \partial^a \phi^{bb} + \TT_{
\frac{1-d}{2}}\phi^{za} + \frac{m\phi^a}{z}\,,
\nonumber\\[5pt]
\tilde\Cbf^z & = & \partial^a \phi^{za} - \half \TT_{\frac{d-1}{2}}\phi^{aa}
+ u \TT_{\frac{3-d}{2}}\phi^{zz} + \frac{m\phi^z}{z}\,,\qquad\quad
\\[5pt]
&& \hspace{-1.2cm} \tilde\Cbf  = \,  \partial^a \phi^a +
\TT_{\frac{3-d}{2}}\phi^z + \frac{m\phi^{aa}}{2z} - \frac{g\phi^{zz}}{(d-2)z}
+ \frac{f\phi}{z}\,,
\nonumber\\[5pt]
&& \hspace{-0.8cm} g \equiv m \Bigl(2\frac{d-2}{d-1}\Bigr)^{1/2}\,,
\qquad
u \equiv   \Bigl(2\frac{d-1}{d-2}\Bigr)^{1/2}\,,
\eeq
where $f$ is defined in \rf{man11112010-20a}. We proceed as follows.

\noindent {\bf i}) First, we note that $\LL_2$ \rf{man13112010-38} can be
represented as
\be \label{man13112010-45} \LL_2  =  \frac{1}{4} \phi^{ab} \Box_\kappa
\phi^{ab} -\frac{1}{8} \phi^{aa} \Box_\kappa \phi^{bb}\,,\ee
where $\kappa$ and $\Box_\kappa$ are given in \rf{man24102010-39} and
\rf{09072009-09} respectively.

\noindent {\bf ii}) Introducing vector fields $\phi_{\pm1}^a$ by the orthogonal
transformation
\beq
&& \phi^{za}  = r_z^{00} \phi_{-1}^a + r_\zeta^{00} \phi_1^a\,,
\nonumber\\[-6pt]
\label{man13112010-46} &&
\\[-6pt]
&& \phi^a  = - r_\zeta^{00} \phi_{-1}^a + r_z^{00} \phi_1^a\,,
\nonumber
\eeq
where $r_z^{00}$, $r_\zeta^{00}$ are given in
\rf{man25102010-01} we cast $\LL_1$ \rf{man13112010-39}
into the form
\be \label{man13112010-48} \LL_1 = \half \sum_{\lambda= \pm 1} \phi_\lambda^a
\Box_{\kappa+\lambda} \phi_\lambda^a\,.\ee
We note that inverse of the transformation
\rf{man13112010-46} is given by
\beq
&& \phi_{-1}^a  = r_z^{00} \phi^{za} - r_\zeta^{00}\phi^a\,,
\nonumber\\[-6pt]
\label{2m27122010-01} &&
\\[-6pt]
&& \phi_1^a  =  r_\zeta^{00} \phi^{za}  + r_z^{00} \phi^a\,.
\nonumber
\eeq

\noindent {\bf iii}) Introducing scalar fields $\phi_0$, $\phi_{\pm 2}$ by
the orthogonal transformation
\beq
&& \phi^{zz}  = s_{11} \phi_{-2} + s_{12}\phi_0 + s_{13}\phi_2 \,,
\nonumber\\[3pt]
\label{man13112010-49} && \phi^z  = s_{21} \phi_{-2} + s_{22}\phi_0 +
s_{23}\phi_2 \,,
\\[3pt]
&& \phi  = s_{31} \phi_{-2} + s_{32}\phi_0 + s_{33}\phi_2 \,,
\nonumber
\eeq
\beq
&& s_{11} = \Bigl(
\frac{(2\kappa+d)(2\kappa+d-2)(d-2)}{16\kappa(\kappa-1)(d-1)}\Bigr)^{1/2}\,,
\nonumber\\[5pt]
&& s_{12} = \Bigl(
\frac{(2\kappa+d)(2\kappa-d)d}{8(\kappa^2-1)(d-1)}\Bigr)^{1/2}\,,
\nonumber\\[5pt]
&& s_{13} = \Bigl( \frac{(2\kappa -d)(2\kappa - d +
2)(d-2)}{16\kappa(\kappa+1)(d-1)}\Bigr)^{1/2}\,,
\nonumber\\[5pt]
&& s_{21} = - \Bigl(
\frac{(2\kappa-d)(2\kappa+d-2)}{8\kappa(\kappa-1)}\Bigr)^{1/2}\,,
\nonumber\\[5pt]
&& s_{22} = \Bigl( \frac{d(d-2)}{4(\kappa^2-1)}\Bigr)^{1/2}\,,
\\[5pt]
&& s_{23} = \Bigl( \frac{(2\kappa+d)(2\kappa - d+
2)}{8\kappa(\kappa+1)}\Bigr)^{1/2}\,,
\nonumber\\[5pt]
&& s_{31} = \Bigl( \frac{(2\kappa -d)(2\kappa - d + 2)d}{16\kappa(\kappa -
1)(d-1)}\Bigr)^{1/2}\,,
\nonumber\\[5pt]
&& s_{32} = - \Bigl( \frac{(2\kappa+d-2)(2\kappa - d + 2)(d-2)}{8(\kappa^2 -
1)(d-1)}\Bigr)^{1/2}\,,\qquad
\nonumber\\[5pt]
&& s_{33} = \Bigl( \frac{(2\kappa+d)(2\kappa + d -
2)d}{16\kappa(\kappa+1)(d-1)}\Bigr)^{1/2}\,,
\nonumber
\eeq
we cast $\LL_0$ \rf{man13112010-40} into the form
\be \label{man13112010-52} \LL_0 = \half \sum_{\lambda=-2,0,2} \phi_\lambda
\Box_{\kappa+\lambda} \phi_\lambda\,.\ee
For the readers convenience, we note that inverse of the transformation
\rf{man13112010-49} is given by
\beq
&& \phi_{-2}  = s_{11} \phi^{zz} + s_{21}\phi^z + s_{31}\phi \,,
\nonumber\\[3pt]
\label{2m27122010-02} && \phi_0  = s_{12} \phi^{zz} + s_{22}\phi^z +
s_{32}\phi \,,
\\[3pt]
&& \phi_2  = s_{13} \phi^{zz} + s_{23}\phi^z + s_{33}\phi \,.
\nonumber
\eeq
\noindent {\bf iv}) Representing $\tilde\Cbf^a$, $\tilde\Cbf^z$, $\tilde\Cbf$
in terms of the vector fields $\phi_{\pm 1}^a$ and the scalar fields
$\phi_0$, $\phi_{\pm 2}$ and introducing $C^a$, $C_{\pm 1}$ by relations
\beq
&& C^a = \tilde\Cbf^a\,,
\nonumber\\[3pt]
&& C_1 = r_\zeta^{00} \tilde\Cbf^z + r_z^{00}
\tilde\Cbf\,,
\\[3pt]
&& C_{-1} = r_z^{00}\tilde\Cbf^z - r_\zeta^{00}
\tilde\Cbf\,,
\nonumber
\eeq
we find that these $C^a$, $C_{\pm1}$ take the form given in
\rf{man05112010-01}. We note the helpful relation

\be  \label{man13112010-56}  \tilde\Cbf^A \tilde\Cbf^A + \tilde\Cbf
\tilde\Cbf = C^aC^a + C_{-1}C_{-1}+ C_1 C_1\,.\ee

\noindent {\bf v}) Making use of relation \rf{man13112010-57}
and taking into account expressions for $\LL_2$ \rf{man13112010-45}, $\LL_1$
\rf{man13112010-48}, $\LL_0$ \rf{man13112010-52} and formula
\rf{man13112010-56},  we see that Lagrangian \rf{man13112010-37} takes the
form of the CFT adapted gauge invariant Lagrangian \rf{man29102010-11}.

We now present some details of the derivation of gauge transformations given
in \rf{man05112010-03a1}. Lagrangian \rf{man11112010-21} is invariant under
gauge transformations given in \rf{man11112010-18}. In terms of canonically
normalized fields \rf{man13112010-28}, these gauge transformations take the
form

\beq
&& \delta \tilde\Phi^{ab} = \partial^a \xi^b + \partial^b \xi^a  -
\frac{2}{z} \eta^{ab}\xi^z + \frac{2m\eta^{ab}}{(d-1)z}\xi \,,\qquad
\nonumber\\[3pt]
&& \delta \tilde\Phi^{za} = \partial^a \xi^z +
\TT_{\frac{d-1}{2}} \xi^a\,,
\nonumber\\[3pt]
&& \delta \tilde\Phi^{zz} = 2\TT_{\frac{d-3}{2}} \xi^z
+ \frac{2m}{(d-1)z}\xi\,,
\nonumber\\[-6pt]
&&
\label{man14112010-07}
\\[-6pt]
&& \delta \tilde\Phi^a = \partial^a \xi - \frac{m}{z}
\xi^a \,,
\nonumber\\[3pt]
&& \delta \tilde\Phi^z = \TT_{\frac{d-3}{2}} \xi -
\frac{m}{z} \xi^z \,,
\nonumber\\[3pt]
&& \delta \tilde\Phi  = - \frac{f}{z} \xi\,.
\nonumber
\eeq
In terms of fields defined in \rf{man13112010-34}, gauge
transformations \rf{man14112010-07} take the form
\beq
&& \hspace{-1.2cm} \delta \phi^{ab} = \partial^a \xi^b
+\partial^b \xi^a + \frac{2\eta^{ab}}{d-2} \TT_{-\frac{d-1}{2}}\xi^z +
\frac{2m\eta^{ab}}{d-2}\xi,
\nonumber\\[5pt]
&& \hspace{-1.2cm}  \delta \phi^{za} = \partial^a
\xi^z + \TT_{\frac{d-1}{2}} \xi^a\,,
\nonumber\\[5pt]
&& \hspace{-1.2cm}  \delta \phi^{zz} = u
\TT_{\frac{d-3}{2}} \xi^z + \frac{m u}{(d-1)z}\xi\,,
\nonumber\\[-6pt]
\label{man14112010-18} &&
\\[-6pt]
&& \hspace{-1.2cm}  \delta \phi^a = \partial^a \xi -
\frac{m}{z} \xi^a \,,
\nonumber\\[5pt]
&& \hspace{-1.2cm}  \delta \phi^z =
\TT_{\frac{d-3}{2}} \xi - \frac{m}{z} \xi^z \,,
\nonumber\\[5pt]
&& \hspace{-1.2cm} \delta \phi  = - \frac{f}{z} \xi\,.
\nonumber
\eeq
Introducing new gauge transformation parameters by the orthogonal
transformation
\beq
&& \xi^z = r_z^{00} \xi_{-1} +  r_\zeta^{00} \xi_1\,,
\nonumber\\[-6pt]
&&
\\[-6pt]
&& \xi = - r_\zeta^{00}\xi_{-1} + r_z^{00} \xi_1\,,
\nonumber
\eeq
and using vector fields $\phi_{\pm1}^a$ \rf{2m27122010-01} and scalar fields
$\phi_0$, $\phi_{\pm 2}$ \rf{2m27122010-02},  we find that gauge
transformations \rf{man14112010-18} take desired form given in
\rf{man05112010-03a1}.

\end{document}